\documentclass[11pt]{article}
\pdfoutput=1 
\usepackage{amssymb}
\usepackage{amsmath}
\usepackage{amstext}
\usepackage{mathrsfs}
\usepackage{multirow}
\usepackage{graphicx}
\usepackage{epsfig}
\usepackage{verbatim} 
\usepackage{fancybox}
\usepackage{color}
\usepackage{mathdots}
\usepackage{anyfontsize}
\usepackage{ulem}
\usepackage{enumitem}
\usepackage{caption}
\usepackage{amsthm}
\usepackage{subfig}
\usepackage{mathtools}
\usepackage{tensor}
\usepackage{slashed}
\usepackage{youngtab}
\usepackage{braket}
\usepackage{bbm}
\usepackage{parskip}
\usepackage[numbers,sort&compress]{natbib}
\usepackage{ytableau}
\usepackage[utf8]{inputenc}
\usepackage{tikz}
\usepackage[compat=1.1.0]{tikz-feynman}
\usepackage{simpler-wick}
\usetikzlibrary{positioning, trees,decorations, decorations.markings, decorations.pathmorphing, arrows, shapes.geometric}

\newcommand{\Comment}[1]{{}}
\definecolor{darkblue}{rgb}{0.15,0.35,0.55}
\definecolor{reddish}{rgb}{0.65, 0.2, 0.2}
\usepackage[linktocpage=true]{hyperref}
\hypersetup{
colorlinks=true,
citecolor=darkblue,
linkcolor=reddish,
urlcolor=darkblue,
pdfauthor={},
pdftitle={},
pdfsubject={}
}

\usepackage{cleveref}

\newcommand{\pvec}[1]{\vec{#1}\mkern2mu\vphantom{#1}}
\newcommand{\be}{\begin{equation}}
\newcommand{\ee}{\end{equation}}

\numberwithin{equation}{section}

\makeatletter
\def\thickhline{%
  \noalign{\ifnum0=`}\fi\hrule \@height \thickarrayrulewidth \futurelet
   \reserved@a\@xthickhline}
\def\@xthickhline{\ifx\reserved@a\thickhline
               \vskip\doublerulesep
               \vskip-\thickarrayrulewidth
             \fi
      \ifnum0=`{\fi}}
\makeatother

\newlength{\thickarrayrulewidth}
\begin{document}

\renewcommand{\thefootnote}{\fnsymbol{footnote}}
~
\vspace{1.75truecm}
\thispagestyle{empty}
\begin{center}
{\LARGE \bf{
Unitarization from Spin Geometry
}}
\end{center}

\vspace{1cm}
\centerline{\begingroup
\Large
	Francesco Bertucci,\footnote{\href{mailto:fbertucc@olemiss.edu}{\texttt{fbertucc@olemiss.edu}}}
\endgroup
\begingroup
\Large
	James Bonifacio,\footnote{\href{mailto:bonifacio@phy.olemiss.edu}{\texttt{bonifacio@phy.olemiss.edu}}}
\endgroup
\begingroup
\Large
and Andrew Schrader\footnote{\href{mailto:awschrad@go.olemiss.edu}{\texttt{awschrad@go.olemiss.edu}}}
\endgroup
}
\vspace{.5cm}

\centerline{{\it Department of Physics and Astronomy,}}
 \centerline{{\it   University of Mississippi, University, MS 38677}} 
 \vspace{.25cm}
 
\vspace{1cm}
\begin{abstract}
\noindent
We derive the dimensional reduction of 10D super Yang--Mills theory on a general closed 6D spin manifold using spectral decompositions for the higher-dimensional gauge field and gaugino, keeping the infinite towers of Kaluza--Klein (KK) modes in the lower-dimensional theory. We then calculate tree-level $2 \rightarrow 2$ scattering amplitudes with external massive spinors and vectors in this lower-dimensional theory. By studying the high-energy behavior of these amplitudes and imposing perturbative unitarity, we derive several sum rules that involve the eigenvalues of Laplace and Dirac operators on the internal 6-manifold, together with products of cubic coupling constants of the KK modes. These give concrete realizations of the known unitarity sum rules for spontaneously broken gauge theories. We illustrate explicitly how the simplest sum rule is satisfied for flat tori and round spheres.

\end{abstract}

\newpage
\setcounter{tocdepth}{3}
\tableofcontents
\renewcommand*{\thefootnote}{\arabic{footnote}}
\setcounter{footnote}{0}

\newpage

\section{Introduction}

Our universe appears to be well described by non-supersymmetric quantum field theory (QFT) and general relativity (GR) in four dimensions. However, this is expected to be just the low-energy effective field theory description that approximates some as-yet-unknown UV complete theory of quantum gravity. The leading paradigm for consistently combining QFT with gravity is string theory, which is naturally formulated as a supersymmetric theory in higher dimensions.  The standard way to reconcile such extra dimensions with observations is through dimensional reduction, an old idea due to Kaluza and Klein \cite{Kaluza:1921tu,Klein:1926tv}, which was initially proposed to unify electromagnetism and gravity. Given a higher-dimensional theory defined on the product of four-dimensional Minkowski spacetime and a small compact internal space, the spectrum and dynamics in 4D can be obtained by integrating over the extra dimensions. While not initially formulated in the context of supersymmetry, this idea has found fruitful applications in supergravity theories (see the classic reviews \cite{Bailin:1987jd,Duff:1986hr}), which in turn can be seen as low-energy descriptions of string theories. Among other features, supersymmetry also provides a mechanism to spontaneously compactify internal dimensions \cite{Freund:1980xh}. Over the decades, numerous attempts have been made to find a realistic Kaluza--Klein (KK) theory, supersymmetric or not, which can suitably describe our 4D world \cite{Witten:1981me,Salam:1981xd,Candelas:1985en}. 

In KK theories, the higher-dimensional fields lead to infinite towers of massive particles in the lower-dimensional theory. If the parent theory contains interactions, these massive fields will be interacting, with coupling constants given by integrals of products of eigenmodes of Laplace and Dirac operators on the internal space, and masses determined by the corresponding eigenvalues. KK theories thus provide a way to construct consistent interacting effective field theories for massive spinning particles, a task which can be difficult if one adopts a bottom-up approach. For example, starting from higher-dimensional Yang--Mills (YM) theories, KK reduction without any truncation leads to interacting theories with infinitely many massive spin-1 particles. These massive spin-1 particles must have $2 \rightarrow 2$ amplitudes that are bounded at high energies, to match the high-energy behavior of higher-dimensional YM theory. The mechanism behind this good high-energy behavior of the KK amplitudes is that the coupling constants and masses in the lower-dimensional theory must satisfy certain sum rules \cite{Chivukula:2001esy, Csaki:2003dt, Ohl:2003dp, Schwinn:2004xa, Schwinn:2005qa, Bonifacio:2019ioc}, which are particular instances of the unitarity sum rules for spontaneously broken gauge theories \cite{LlewellynSmith:1973yud, Cornwall:1973tb, Gunion:1990kf}.

Similarly, starting from pure GR in higher dimensions, KK reduction without any truncation defines a theory with an infinite tower of massive spin-2 particles, as well as massive scalars and massive vectors, the spectrum of which is described for a general closed internal manifold in \cite{Hinterbichler:2013kwa}. The scattering amplitudes of these interacting massive spin-2 particles must inherit the relatively good high-energy behavior of graviton amplitudes in higher-dimensional GR \cite{SekharChivukula:2019yul, Chivukula:2020hvi, Hang:2021fmp, Bonifacio:2019ioc}, which is much better than is possible for theories with only finitely many healthy massive spin-2 particles \cite{Arkani-Hamed:2002bjr, Schwartz:2003vj, deRham:2010ik,Bonifacio:2019mgk, Bonifacio:2019ioc}. As with spin-1 particles, to achieve this improved high-energy behavior, the coupling constants and masses must satisfy sum rules \cite{Bonifacio:2019ioc, SekharChivukula:2019qih,Chivukula:2022kju, deGiorgi:2023mdy}. From a geometrical perspective, the sum rules are consistency conditions that must be satisfied by the eigenvalues and triple integrals of eigenmodes on the internal space.

For closed internal manifolds, the sum rules can also be obtained directly by considering quadruple integrals of eigenmodes on the internal manifold and expanding them in different ways using associativity and completeness of the eigenmodes \cite{Bonifacio:2019ioc}.
This is analogous to how one obtains conformal field theory (CFT) consistency conditions, which underlie the conformal bootstrap \cite{Belavin:1984vu,Rattazzi:2008pe}. Following the approach of the numerical conformal bootstrap \cite{Poland:2018epd}, this allows for semi-definite optimization to be used to derive bounds on eigenvalues and triple products on manifolds and orbifolds. This idea has been explored in a variety of contexts \cite{Bonifacio:2020xoc, Bonifacio:2021msa,Bonifacio:2021aqf, Kravchuk:2021akc, Radcliffe:2024jcg, Bonifacio:2023ban, Gesteau:2023brw, Adve:2025rvf, Bertucci:2026fjh}. In particular, the case of the Dirac operator on hyperbolic spin surfaces was considered in \cite{Gesteau:2023brw}, giving upper bounds on the Dirac spectral gap. For hyperbolic manifolds, the analogy with CFTs can be made precise due to the underlying conformal group \cite{Kravchuk:2021akc}. 

The goal of this paper is to push this program forward, by initiating a study of higher-dimensional spin manifolds and supersymmetric theories. In particular, we study KK scattering amplitudes coming from supersymmetric Yang--Mills (SYM) theory in 10 dimensions \cite{Gliozzi:1976qd}. This theory and its lower-dimensional descendants have been extremely well-studied and provide a useful testbed for this approach. We start by compactifying 10D SYM down to four dimensions over an arbitrary closed 6-dimensional spin manifold. This requires decomposing the 10D fields using harmonic decompositions for spinors and vectors. We then calculate scattering amplitudes in the lower-dimensional theory, keeping the infinite towers of massive KK modes. The tree-level $2 \rightarrow 2$ amplitudes in 10D SYM are bounded at high energies, so this should also hold for amplitudes in the dimensionally reduced theory. However, the amplitudes with external massive KK modes naively grow at high energies. Cancellation of the would-be bad high-energy behavior leads to sum rules involving the masses and cubic coupling constants of the KK modes, which are determined by eigenvalues and triple products of eigenmodes of Laplace and Dirac operators on the six-dimensional spin manifold. We show how to derive the sum rules directly by using associativity and completeness to write quadruple integrals of eigenmodes on the internal space in different ways  and we explicitly verify a Dirac spinor version of the simplest sum rule for flat tori and round spheres. These sum rules are particular instances (with infinitely many fields) of the general unitarity sum rules for spontaneously broken gauge theories with fermions \cite{LlewellynSmith:1973yud, Gunion:1990kf}. Some instances of these sum rules have also been studied in the context of 5D KK theories with fermions \cite{Schwinn:2004xa, Schwinn:2005qa}. 

The rest of this paper is organized as follows: In Section \ref{sec:2}, we perform the dimensional reduction of 10D SYM down to 4D Minkowski spacetime, where the internal space is a general 6-dimensional spin manifold that is orientable and closed, i.e., compact without boundary. We also give the gauge and supersymmetry (SUSY) transformations in the lower-dimensional theory. In Section \ref{sec:3}, we compute tree-level $2 \rightarrow 2$ scattering amplitudes involving the massive KK modes and take their high-energy limits to obtain sum rules. Our conclusions are given in Section \ref{sec:conclusions}. Appendix \ref{app:clifford} contains our gamma matrix conventions and Appendix \ref{app:examples} shows explicitly how the simplest sum rule is satisfied for flat tori and round spheres.

\paragraph{Notation and conventions.}

We mostly follow the notation of Polchinski's book \cite{Polchinski_1998}. In particular, the signature of the metric is $(-,+,+,\dots,+)$. The Ricci tensor and Ricci scalar are denoted by $R_{MN}$ and $R$ in 10D and by $R^{(6)}_{pq}$ and $R^{(6)}$ in 6D. We (anti)symmetrize with weight 1. The number of supersymmetries is $\mathcal{N}=1$ in 10D and $\mathcal{N}_{4}\in\{0,1,2,4\}$ in 4D.

\section{Action and Kaluza--Klein reduction}\label{sec:2}

\subsection{10D SYM}

Let $\mathbb{R}^{1,3}\times \mathcal{M}$ be a Lorentzian manifold in $D=10$ dimensions given by the product of four-dimensional Minkowski spacetime and a six-dimensional internal space $\mathcal{M}$, where $\mathcal{M}$ is a closed, connected, orientable Riemannian spin manifold.\footnote{Recall that an orientable Riemannian manifold $\mathcal{M}$ is spin if and only if its second Stiefel–Whitney class vanishes, $w_2(\mathcal{M}) =0$.
If this is the case, the possible spin structures are in one-to-one correspondence with the elements of the first cohomology group $H^1(\mathcal{M},\mathbb{Z}_2)$. Simply connected spin manifolds, such as a K3 surface, have a unique spin structure, namely, the trivial one. A flat $n$-torus has $2^n$ possible spin structures, corresponding to the choice of periodic or antiperiodic boundary conditions in each direction.}

We consider $D=10$, $\mathcal{N}=1$ SYM theory on $\mathbb{R}^{1,3}\times \mathcal{M}$ and perform dimensional reduction down to $d=4$ dimensions. The higher-dimensional metric is
\begin{equation}
	ds^2 =G_{MN}dX^M dX^N =\eta_{\mu\nu}dx^\mu dx^\nu +g_{pq}dy^p dy^q,
\end{equation}
where spacetime indices in 10D are $M,N,\dots$ and go from 0 to 9, indices in 4D are $\mu,\nu,\dots$ and go from 0 to $3$, and indices in 6D are $p,q,\dots$ and go from 1 to $6$. The four-dimensional Minkowski metric is denoted by $\eta_{\mu\nu}$ and the curved metric on $\mathcal{M}$ is denoted by $g_{pq}$. Coordinates in ten, four and six dimensions are denoted by $X$, $x$, and $y$, respectively. 

Using $A,B,\dots$ to denote color indices, the field content is given by a massless vector $V_M^A(X)$ and a Majorana--Weyl (MW) spinor $\lambda^A(X)$, which are both in the adjoint representation of a compact gauge group $G$.
The 10D MW conditions are given by
\begin{equation}\label{eq:MWcond}
	\begin{split}
		\big(\lambda^A\big)^* &=B\lambda^A, \\
		P_-\lambda^A &=\lambda^A,
	\end{split}
\end{equation}
where
\begin{equation} \label{eq:10D_B}
	B=\Gamma^3\Gamma^5\Gamma^7\Gamma^9
\end{equation}
and the chirality projectors are given by
\begin{equation}
	P_\pm \coloneq \frac{I_{10}\pm\Gamma}2,
\end{equation}
with $I_{10}$ the identity matrix and $\Gamma$ the chirality matrix in ten dimensions. 

The 10D SYM action is \cite{Gliozzi:1976qd}
\begin{equation}\label{eq:SYM10dlagr2}
	S=S_{\rm YM}+S_{\rm D}= \int d^{10}X \sqrt{-G} \left[ -\frac14 F^A_{MN}F^{MN}_A -\frac{i}{2} \bar\lambda^A \Gamma^M (D_M \lambda)_A \right]  ,
\end{equation}
where the field strength and covariant derivative are given by
\begin{equation}
	\begin{split}
		F^A_{MN} &= \partial_M V^A_N -\partial_N V^A_M +g {f^A}_{BC} V^B_M V^C_N, \\
		(D_M \lambda)^A &= \nabla_M \lambda^A +g {f^A}_{BC} V^B_M \lambda^C,
	\end{split}
\end{equation}
where $g$ is the YM coupling in ten dimensions, ${f^A}_{BC}$ are the structure constants for $G$ and $\nabla$ is the Levi-Civita connection. Repeated spacetime and color indices are always summed over.

The action \eqref{eq:SYM10dlagr2} is invariant under the gauge transformations
\begin{equation}\label{eq:10dGaugeTransf}
	\begin{split}
		\delta V^A_M &=\partial_M \Lambda^A +g{f^{A}}_{BC} V^B_M \Lambda^C, \\
		\delta \lambda^A &=g{f^{A}}_{BC} \lambda^B \Lambda^C,
	\end{split}
\end{equation}
where $\Lambda^A(X)$ are 10D scalar gauge parameters. The action \eqref{eq:SYM10dlagr2} is also invariant under the $\mathcal{N}=1$ supersymmetry transformations
\begin{equation}\label{eq:SUSYtransf10d}
	\begin{split}
		\delta_{\rm SUSY} V^A_M &=-i\bar\xi \Gamma_M \lambda^A, \\
		\delta_{\rm SUSY} \lambda^A &=\frac12 F^A_{MN} \Gamma^{MN}\xi,
	\end{split}
\end{equation}
where $\xi(X)$ is a covariantly constant 10D MW spinor parameter and $\Gamma^{MN\cdots Q} \coloneq \Gamma^{[M} \Gamma^N \cdots \Gamma^{Q]}$.

\subsection{Field decompositions}
 
We are interested in the following decomposition of the 10D spin group:
\begin{equation}
	{\rm Spin}(1,9) \, \longrightarrow \, {\rm Spin}(1,3) \times {\rm Spin}(6).
\end{equation}
Under restriction to this subgroup, a ten-dimensional negative chirality spinor of ${\rm Spin}(1,9)$ decomposes as $\boldsymbol{16}\to(\boldsymbol{2},\boldsymbol{4})\oplus(\bar{\boldsymbol{2}},\bar{\boldsymbol{4}})$. Note that $\boldsymbol{2},\bar{\boldsymbol{2}}$ are Spin$(1,3) \cong { SL}(2, \mathbb{C})$ Weyl representations in 4D, and  $\bar{\boldsymbol{2}}$ is the complex conjugate of $\boldsymbol{2}$. Similarly, $\boldsymbol{4},\bar{\boldsymbol{4}}$ are Spin$(6) \cong  SU(4)$  Weyl representations in 6D, and  $\bar{\boldsymbol{4}}$ is the complex conjugate of $\boldsymbol{4}$.

We decompose 10D indices as $M=(\mu,p)$, with $\mu=0,\dots,3$ and $p=1,\dots,6$. We choose a representation for the 10D gamma matrices $\Gamma^M$ such that
\begin{equation}\label{eq:gammamatrixtext}
	\begin{split}
		\Gamma^\mu &=\gamma^\mu \otimes \tilde\Gamma, \qquad \mu=0,\dots,3, \\
		\Gamma^p &=I_4 \otimes \tilde\Gamma^{p}, \qquad p=1,\dots,6,
	\end{split}
\end{equation}
where $\gamma^\mu$ are gamma matrices in four dimensions (satisfying the usual Lorentzian Clifford algebra), $\tilde\Gamma^p$ are gamma matrices in six dimensions (satisfying the Euclidean Clifford algebra), and  $\tilde\Gamma$  is the chirality matrix in six dimensions. Details on the construction, as well as explicit forms of these matrices, can be found in Appendix \ref{app:clifford}. The 10D chirality matrix and the matrix $B$, defined in \eqref{eq:10D_B}, become
\begin{align}
\Gamma &=-\gamma_5\otimes\tilde\Gamma, \\
B &=\gamma_5 B_4\otimes\tilde\Gamma C_6, 
\end{align}
where $B_4 =\gamma_5\gamma^3$ and	$C_6 =\tilde\Gamma^2\tilde\Gamma^4\tilde\Gamma^6$. Here $\gamma_5$ is the chirality matrix in four dimensions and $B_4$ is used to define a Majorana spinor in four dimensions (see \eqref{eq:4DMajorana} below).

\subsubsection{Chiral spinors} \label{sec:spinors}

Let $\chi_{\pm,i}(y)$ be spinors on $\mathcal{M}$ of definite chirality, $\tilde\Gamma\chi_{\pm,i} =\pm\chi_{\pm,i}$, that satisfy
\begin{subequations}
	\begin{align}
		i\tilde\Gamma^p\nabla_p\chi_{\pm,i} &=\lambda_i\chi_{\mp,i}, \label{eq:6dDiracEigenprob} \\
		\left(\chi_{+,i}\right)^* &=C_6\chi_{-,i}, \label{eq:6dDiracConjugate} \\
		\int_\mathcal{M} \chi_{s,j}^\dagger \chi_{s',i} &=\delta_{ss'}\delta_{ij}, \label{eq:6dDiracEigenprobON}
	\end{align}
\end{subequations}
where $\lambda_i$ are real numbers and integrals over the internal manifold are written as $\int_\mathcal{M} \coloneq \int_\mathcal{M} d^6y \sqrt{g}$. These form bases of square-integrable spinors on the internal manifold $\mathcal{M}$ with definite chirality. We can take $\lambda_i \geq 0$ without loss of generality, since $\chi_{\pm, i} \rightarrow \pm i \chi_{\pm, i}$ sends $\lambda_i \rightarrow - \lambda_i$. By \eqref{eq:6dDiracEigenprob}, the Dirac operator flips the chirality of spinors with non-zero eigenvalues, so $\chi_{+, i} \pm \chi_{-,i}$ are eigenspinors of the Dirac operator with eigenvalue $\pm \lambda_i$.  The condition \eqref{eq:6dDiracConjugate} reflects that for Euclidean spinors in $n=6$, the left and right representations are complex conjugates of each other.

The zero modes of the six-dimensional Dirac operator are called harmonic spinors, which we can take to have definite chirality. We label them using the index $\alpha$, so that $i\tilde\Gamma^p\nabla_p\chi_{\pm,\alpha} =0$ (unless stated otherwise, the index $i$ also includes the harmonic spinors). 
A geometrical constraint for the existence of harmonic spinors is given by the Lichnerowicz–Weitzenb\"ock
formula \cite{Lichnerowicz:1963},
\begin{equation}\label{eq:Lich}
	(i\tilde\Gamma^p\nabla_p)^2 =-\nabla^p\nabla_p +\frac{R^{(6)}}4,
\end{equation}
where $R^{(6)}$ is the Ricci scalar on $\mathcal{M}$. Since the Laplacian $-\nabla^p\nabla_p$ is non-negative, spin manifolds cannot admit harmonic spinors if $R^{(6)}$ is everywhere non-negative and not identically zero.

The choice of spin structure affects the form of the Dirac spectrum and, in particular, the number of harmonic spinors $\chi_{\pm,\alpha}$ (see, e.g., Ref.~\cite{baer2000}).
However, by the Atiyah--Singer index theorem \cite{Atiyah:1963zz}, the index of  $i\tilde\Gamma^p\nabla_p$,
\begin{equation} \label{eq:index}
	\mathrm{index}(i\tilde\Gamma^p\nabla_p) \coloneq n^{(6)}_+ - n^{(6)}_-,
\end{equation}
is a topological invariant of $\mathcal{M}$ and therefore does not depend on the choice of spin structure. Here $n^{(6)}_+$ and $n^{(6)}_-$ denote the number of harmonic spinors with positive and negative chirality, respectively. In six dimensions, the index \eqref{eq:index} vanishes and thus $n^{(6)}_+ = n^{(6)}_-$ \cite{Eguchi:1980jx,Green_Schwarz_Witten_2012,Witten:1983ux,Witten:1981me}. For non-trivial gauge field backgrounds, the index of the twisted Dirac operator will in general depend on the gauge field expectation value and can be non-zero. For an explicit expression of the index in six dimensions, see \cite{Green_Schwarz_Witten_2012}.

\subsubsection{Spinor decomposition}

Using the chiral spinors $\chi_{\pm, i}$, the Kaluza--Klein (KK) decomposition for the ten-dimensional MW spinors $\lambda^A(X)$ can be written as
\begin{equation}\label{eq:KKspinor}
	\lambda^A =\sum_i \left[ \psi^A_{L,i}(x) \otimes \chi_{+,i}(y) +\psi^A_{R,i}(x) \otimes \chi_{-,i}(y) \right],
\end{equation}
where the 4D spinors satisfy
\be 
\left(\psi_{L,i}^A\right)^* =B_4\psi_{R,i}^A, \qquad \gamma_5\psi_{L,i}^A =\psi_{L,i}^A, \qquad		\gamma_5\psi_{R,i}^A =-\psi_{R,i}^A.
\ee
With these constraints, the ansatz \eqref{eq:KKspinor} automatically satisfies the 10D MW conditions \eqref{eq:MWcond}. 

It is convenient to define the spinors
\begin{equation}
	\psi^A_i \coloneq\psi^A_{L,i} +\psi^A_{R,i},
\end{equation}
which are $d=4$ Majorana spinors, i.e.,
\begin{equation}\label{eq:4DMajorana}
	\left(\psi^A_i\right)^* =B_4 \psi^A_i.
\end{equation}
The left and right components can be obtained with the $d=4$ projectors,
\begin{equation}
		P_L =\frac{I_4+\gamma_5}2, \qquad		P_R =\frac{I_4-\gamma_5}2.
\end{equation}

\subsubsection{Vector decomposition}
We denote by $\zeta_a(y)$ the non-constant Laplace eigenfunctions on $\mathcal{M}$, 
\be
-\nabla_p\nabla^p \zeta_a =\lambda_a \zeta_a,
\ee
where $\lambda_a>0$. We take these to be real and orthonormal,
\begin{equation}
(\zeta_a)^*=\zeta_a, \qquad \int_\mathcal{M} \zeta_a\zeta_b =\delta_{ab}.
\end{equation}
We treat separately the normalized zero-mode, $\zeta_0 =1/\sqrt{V}$, where $V$ is the volume of $\mathcal{M}$. Taken together, these eigenfunctions form an orthonormal basis for square-integrable functions on $\mathcal{M}$.

Similarly, we denote by $Y_{p,I}(y)$ the divergence-free eigenvectors of the Hodge Laplacian on $\mathcal{M}$, 
\begin{equation}
	-\nabla_q\nabla^q Y_{p,I} +{{R^{(6)}}_p}^q Y_{q,I} =\lambda_I Y_{p,I}, \quad\quad \nabla^p Y_{p,I} =0,
\end{equation}
where $\lambda_I \geqslant 0$ and $R^{(6)}_{pq}$ is the Ricci tensor on $\mathcal{M}$. We take these to be real and orthonormal,
\be
(Y^p_I)^*=Y^p_I, \qquad \int_\mathcal{M} Y^p_I Y_{p,J} =\delta_{IJ}.
\ee
These form a real orthonormal basis of square-integrable divergence-free vectors on the internal manifold $\mathcal{M}$.

We can combine a vector Hodge decomposition on $\mathcal{M}$ with spectral expansions using the above eigenmodes to write the gauge fields $V^A_M(X)$ as
\begin{equation}\label{eq:KKvector}
	V^A_M =\begin{pmatrix}
		V^A_\mu(x,y) \\ V^A_p(x,y)
	\end{pmatrix} =\begin{pmatrix}
		\sum_a A^{Aa}_\mu(x) \zeta_a(y) +\frac1{\sqrt{V}} A^{A0}_\mu(x) \\
		\sum_a A^{Aa}(x) \nabla_p\zeta_a(y) +\sum_I \phi^{AI}(x)Y_{p,I}(y)
	\end{pmatrix}.
\end{equation}

\subsection{Gauge fixing}

The $x$-dependent coefficients in the decompositions \eqref{eq:KKspinor} and \eqref{eq:KKvector} correspond to 4D fields in the KK reduced theory: $A^{A0}_\mu $ and $A^{Aa}_\mu$ are vectors, $\psi^A_{L,i}$ and $\psi^A_{R,i}$ are spinors, and $A^{Aa} $ and $\phi^{AI} $ are scalars. We can see how these fields transform under gauge transformations by decomposing the gauge parameter,
\begin{equation}
	\Lambda^A =\sum_a \Lambda^{Aa}(x) \zeta_a(y) +\frac1{\sqrt{V}} \Lambda^{A0}(x).
\end{equation}
The gauge transformations in 10D \eqref{eq:10dGaugeTransf} then imply the following gauge transformations of the 4D fields:
\begin{equation}
	\begin{split}
		\delta A^{A0}_\mu &=\partial_\mu \Lambda^{A0} +\frac{g}{\sqrt{V}}{f^A}_{BC}A^{B0}_\mu \Lambda^{C0}  +\frac{g}{\sqrt{V}}{f^A}_{BC}\sum_a A^{Ba}_\mu \Lambda^{Ca}, \\
		\delta A^{Aa}_\mu &=\partial_\mu \Lambda^{Aa} +\frac{g}{\sqrt{V}}{f^A}_{BC} \left(A^{B0}_\mu \Lambda^{Ca} +A^{Ba}_\mu \Lambda^{C0} \right) +g{f^A}_{BC} \sum_{b,c} c_{abc} \, A^{Bb}_\mu \Lambda^{Cc}, \\
		\delta A^{Aa} &= \Lambda^{Aa} +\frac{g}{\sqrt{V}}{f^A}_{BC}A^{Ba} \Lambda^{C0} \\
		&\quad +g{f^A}_{BC} \sum_{b,c} \frac12 c_{abc} \frac{\lambda_a+\lambda_b -\lambda_c}{\lambda_a} \, A^{Bb}\Lambda^{Cc} +g{f^A}_{BC} \sum_{I,c} \frac{c_{acI}}{\lambda_a} \, \phi^{BI}\Lambda^{Cc}, \\
		\delta \phi^{AI} &=\frac{g}{\sqrt{V}}{f^A}_{BC}\phi^{BI} \Lambda^{C0} +g{f^A}_{BC} \sum_{b,c} c_{bcI} \,A^{Bb}\Lambda^{Cc} +g{f^A}_{BC} \sum_{J,c} c_{cIJ} \, \phi^{BJ}\Lambda^{Cc}, \\
		\delta\psi^A_{L,i} &=\frac{g}{\sqrt{V}}{f^A}_{BC} \, \psi^B_{L,i} \Lambda^{C0} +g{f^A}_{BC} \sum_{j,a} c_{ija}\, \psi^{Bj}_{L} \Lambda^{Ca},
	\end{split}
\end{equation}
where we have defined the real triple overlaps
\begin{equation}
	\begin{split}
		c_{abc} &\coloneq\int_\mathcal{M} \zeta_a\zeta_b\zeta_c, \\
		c_{abI} &\coloneq\int_\mathcal{M} Y^q_I \, \nabla_q \zeta_a \, \zeta_b, \\
		c_{aIJ} &\coloneq\int_\mathcal{M} \zeta_a \, Y^p_J \, Y_{p,I},
	\end{split}
\end{equation}
and the complex triple overlaps
\begin{equation}\label{eq:tripleoverlapFermion1}
	c_{jia} \coloneq \int_\mathcal{M} \zeta_a\chi_{+,j}^\dagger\chi_{+,i}.
\end{equation}
Note that $c_{abc}$ is symmetric under any permutation of its indices, whereas $c_{abI}=-c_{baI}$, $c_{aIJ}=c_{aJI}$, and $(c_{jia})^*=c_{ija}$.

We can partially fix the gauge symmetry by perturbatively fixing $\Lambda^{Aa}$ order-by-order in $g$ to reach the unitary gauge
\begin{equation}\label{eq:YMgauge}
	A^{Aa}=0.
\end{equation}
The residual gauge transformations come from the zero mode $\Lambda^{A0}$ of the gauge parameter and read
\begin{equation}
	\begin{split}
		\delta A^{A0}_\mu &=\partial_\mu \Lambda^{A0} +\frac{g}{\sqrt{V}}{f^A}_{BC}A^{B0}_\mu \Lambda^{C0}, \\
		\delta A^{Aa}_\mu &=\frac{g}{\sqrt{V}}{f^A}_{BC} A^{Ba}_\mu \Lambda^{C0}, \\ \delta\phi^{AI} &=\frac{g}{\sqrt{V}}{f^A}_{BC}\phi^{BI} \Lambda^{C0}, \\
		\delta\psi^A_{L,i} &=\frac{g}{\sqrt{V}}{f^A}_{BC} \, \psi^B_{L,i} \Lambda^{C0}.
	\end{split}
\end{equation}
The four-dimensional fields thus transform in the adjoint representation of the gauge group.

\subsection{4D action}

We can plug the decompositions \eqref{eq:KKspinor} and \eqref{eq:KKvector} into the action \eqref{eq:SYM10dlagr2} and integrate over the internal manifold $\mathcal{M}$ to get a four-dimensional action. If we further perform the chirality rotation
\begin{equation} \label{eq:phase_rotation}
	\psi^A_i \,\,\to\,\, e^{i\frac\pi4\gamma_5}\psi^A_i =\frac1{\sqrt{2}}(I_4 +i\gamma_5)\psi^A_i,
\end{equation}
the Dirac part of the resulting action in unitary gauge \eqref{eq:YMgauge} is
\begin{equation}\label{eq:4dFinalDirac}
	\begin{split}
		S_{\rm D} &=\int d^4x \bigg[ -\frac12\sum_i \bar\psi_{iA} \left(i\gamma^\mu \nabla_\mu -i\lambda_i\right)\psi^{Ai} -\frac{i}2 \frac{g}{\sqrt{V}} f_{ABC} \sum_i A^{B0}_\mu \bar\psi^A_i \gamma^\mu \psi^{Ci} \\
		&\quad\quad\quad\quad\quad -\frac{i}2gf_{ABC}\sum_{i,j,a}  A^{Ba}_\mu \bar\psi^{Aj} \gamma^\mu\Big(c_{jia} P_L +\left(c_{jia}\right)^*P_R \Big) \psi^{Ci} \\
		&\quad\quad\quad\quad\quad\quad +\frac12gf_{ABC}\sum_{i,j,I} \phi^{BI} \bar\psi^{Aj}\Big(-c_{jiI} P_R +(c_{jiI})^* P_L\Big)\psi^{Ci} \bigg],
	\end{split}
\end{equation}
where we have defined the triple overlaps
\begin{equation}\label{eq:tripleoverlapFermion2}
	c_{jiI} \coloneq \int_\mathcal{M} Y_{p,I} \chi_{+,j}^\dagger \tilde\Gamma^{p} \chi_{-,i},
\end{equation}
which satisfy $c_{jiI}=-c_{ijI}$.
The fermionic degrees of freedom described by  \eqref{eq:4dFinalDirac} consist of a tower of massive Majorana spinor multiplets $\psi^A_i$ with masses $m_i=|\lambda_i|$, including a massless spinor multiplet for each pair of harmonic spinors $\chi_{\pm, \alpha}$.

The four-dimensional YM action $S_{\rm YM}$ in the unitary gauge \eqref{eq:YMgauge} is written explicitly in \cite{Bonifacio:2019ioc}. The spectrum consists of a single multiplet of massless vectors $A^{A0}_\mu$, a tower of massive vector multiplets $A^{Aa}_\mu$ with squared masses $m^2_a=\lambda_a$, a tower of scalar multiplets $\phi^{AI}$ with squared masses $m_I^2=\lambda_I$, including a massless scalar multiplet for each harmonic vector.

Taking the zero-mode truncation of the lower-dimensional SYM theory on a six-dimensional flat torus $\mathcal{M}=T^6$, equipped with the trivial spin structure, gives the conformally invariant $d=4$, $\mathcal{N}_4=4$ SYM theory.\footnote{The full theory including the massive modes does not have conformal symmetry.} In this theory, the massless real scalars $\phi^{BI_0}$ ($I_0=1,\dots,6$) are in the vector representation $\mathbf{6}$ of the global internal $R$-symmetry group $SU(4)\cong\mathrm{Spin}(6)$, while the massless fermions $\psi^A_{\alpha}$ ($\alpha=1,\dots,4$) sit in the fundamental representation $\mathbf{4}$. The triple overlaps $c_{\beta\alpha I_0}$ are the Yukawa couplings in $d=4$, while they represent the Clebsch–Gordan coefficients which intertwine the representation $\mathbf{6}$ with $\mathbf{4}$, $\bar{\mathbf{4}}$ in six dimensions. These couplings are proportional to the Euclidean gamma matrices in $n=6$. Indeed, using the results of Appendix \ref{sec:exampletori}, we have
\begin{equation}
	{c_{\beta\alpha}}^{I_0}= \big(\tilde\Gamma^{I_0}\big)_{\beta\alpha}/\sqrt{V}.
\end{equation}
\subsection{SUSY transformations}

We now discuss the form of the SUSY transformations in the 4D theory before fixing unitary gauge.  

\subsubsection{Parallel spinors}

The invariance under global supersymmetry transformations of the SYM action \eqref{eq:SYM10dlagr2} on flat spacetime was proved in \cite{Gliozzi:1976qd,Brink:1976bc}. In general, when a theory is put on a curved background, the SUSY spinor parameter acquires a spacetime dependence. The difficulty of defining a supersymmetric theory on curved space is well known, see for instance \cite{Blau:2000xg,Festuccia:2011ws}.
The theory \eqref{eq:SYM10dlagr2} can be defined on a curved manifold and be SUSY invariant provided that the SUSY spinor parameter $\xi(X)$ is covariantly constant, i.e.,
\begin{equation}\label{eq:nablaxi0}
	\nabla_M\xi =0.
\end{equation}

Since $\xi(X)$ is a 10D MW spinor, it can be decomposed on the space $\mathbb{R}^{1,3}\times\mathcal{M}$ as 
\begin{equation}\label{eq:xiSUSYpre}
	\xi =\sum_i \Big[\xi_{L,i}(x)\otimes\chi_{+,i}(y) +\xi_{R,i}(x)\otimes\chi_{-,i}(y)\Big],
\end{equation}
for some $d=4$ SUSY spinor parameters $\xi_{L,i} = (B_4\xi_{R,i})^*$.
The condition \eqref{eq:nablaxi0} then reduces to
\begin{equation}\label{eq:covconstSpin}
	\begin{split}
		\partial_\mu \xi_{L,i} &=\partial_\mu \xi_{R,i} =0, \\
		\sum_i \xi_{L,i}\otimes\nabla_p\chi_{+,i}(y) &=\sum_i \xi_{R,i}\otimes\nabla_p\chi_{-,i}(y) =0.
	\end{split}
\end{equation}
By combining these two equations, we can conclude that SUSY invariance in $D=10$ implies
\begin{equation}\label{eq:covconstSpin2}
	\begin{split}
		\xi_{L,\alpha} &=\big(B_4\xi_{R,\alpha}\big)^* = \mathrm{const}, \\
		\xi_{L,i} &=\xi_{R,i} =0, \quad\quad \textrm{if} \quad\quad \lambda_i\ne0, \\
		\nabla_p\chi_{\pm,\alpha} &=0, \quad\quad \textrm{if} \quad\quad \xi_{L,\alpha}= \big(B_4\xi_{R,\alpha}\big)^* \ne0,
	\end{split}
\end{equation}
where we recall that $\alpha$ labels the harmonic spinors on $\mathcal{M}$. Thus, in \eqref{eq:xiSUSYpre} we need to sum over the harmonic spinors $\chi_{\pm,\alpha}$ that are also parallel:
\begin{equation}\label{eq:xiSUSYparam_precursor}
	\xi =\sum_{\substack{\alpha, \\\textrm{parallel}}} \Big[\xi_{L,\alpha}\otimes\chi_{+,\alpha}(y)+\xi_{R,\alpha}\otimes\chi_{-,\alpha}(y)\Big].
\end{equation}
The maximal number of unbroken supersymmetries $\mathcal{N}_{4}$ in the $d=4$ theory is thus determined by the number of parallel spinors of each chirality on the internal manifold \cite{Witten:1981me,Duff:1982ev}.

An upper bound on the number of independent parallel spinors on the internal spin manifold $\mathcal{M}$ is determined by the restricted holonomy group $H_0\subseteq \mathrm{Spin}(6)\cong SU(4)$ of its connection \cite{Awada:1982pk,Duff:1983vj,Duff:1983gq,Duff:1986hr,Duff:2025tot,Bailin:1987jd}. 
For a six-dimensional internal spin manifold $\mathcal{M}$, there are four possibilities:
\begin{itemize}
	\item The holonomy satisfies $H_0\not\subseteq SU(3)$ for any $SU(3)\subset SU(4)$. In this case, there are no parallel spinors. The internal manifold $\mathcal{M}$ is a generic six-dimensional curved manifold. All supersymmetry is broken and the KK reduction will yield a non-supersymmetric $d=4$ theory. This theory can still have massless fermions if the internal manifold admits harmonic spinors.
	\item The holonomy satisfies $H_0\subseteq SU(3)$ but $H_0\not\subseteq SU(2)$. In this case, there is at most one parallel spinor of each chirality and the internal manifold $\mathcal{M}$ is a Calabi--Yau threefold \cite{Candelas:1985en}. Supersymmetry may be partially broken in $d=4$, with $\mathcal{N}_{4}\le1$.
	\item The holonomy satisfies $H_0\subseteq SU(2)$ with $H_0\ne 1$ (identity group).  In this case, there can be at most  two parallel spinors of each chirality. The internal manifold $\mathcal{M}$ is (finitely covered by) the product of a K3 surface and a two-dimensional flat torus, i.e., $\mathrm{K3}\times T^2$ \cite{Duff:1983vj}. Supersymmetry may be partially broken in $d=4$, with $\mathcal{N}_{4}\le2$.
	\item The holonomy is trivial, $H_0=1$. In this case, there can be at most four parallel spinors of each chirality. The internal manifold $\mathcal{M}$ is a six-dimensional flat manifold. Supersymmetry may be unbroken in $d=4$, with $\mathcal{N}_{4}\le4$.
\end{itemize}
We see that parallel spinors are relatively rare, and their presence puts stringent constraints on the geometry of the internal spin manifold $\mathcal{M}$. In particular, manifolds with $H_0\subseteq SU(3)$  are Ricci flat \cite{HITCHIN19741,Gubser:2002mz},
\begin{equation}\label{eq:6dRicciFlat}
	R^{(6)}_{pq}=0.
\end{equation}
For Ricci-flat manifolds, all harmonic spinors are parallel, and therefore we can simply write \eqref{eq:xiSUSYparam_precursor} as
\begin{equation} \label{eq:xiSUSYparam}
	\xi =\sum_\alpha \Big[\xi_{L,\alpha}\otimes\chi_{+,\alpha}(y)+\xi_{R,\alpha}\otimes\chi_{-,\alpha}(y)\Big].
\end{equation}

Within these possibilities, the choice of spin structure determines the actual number of supersymmetries in $d=4$. Let us mention some important cases: on a flat torus, the parallel spinors are constant, and thus they can be non-zero if and only if the spin structure is the trivial one. Similarly, since $\mathrm{K3}$ is simply connected, for the space $\mathrm{K3}\times T^2$ the same conclusion applies. If $\mathcal{M}$ is a simply connected Calabi--Yau threefold, there is only the trivial spin structure. In physics terms, in all these cases the boundary conditions for the spinors must be periodic (or of Ramond type) in all directions. The last case we discuss is non-simply connected Calabi--Yau threefolds. For these, spin structures with parallel spinors are in one-to-one correspondence with lifts to $\mathrm{Spin}(6)$ of the (global) holonomy group $H$, with fixed points on the spin representation space \cite{10.1063/1.532293,Wang1989,Wang1995,Moroianu_2000}. When the spin structure is not of this type, supersymmetry is totally broken even though $H_0\subseteq SU(3)$ and $H_0\not\subseteq SU(2)$.

\subsubsection{4D SUSY}

Let us now work out how the 10D SUSY transformations are realized on 4D fields. The $D=10$ supersymmetry transformations \eqref{eq:SUSYtransf10d} read
\begin{subequations}
	\label{eq:4dSUSYtransf}
	\begin{align}
		\delta_{\rm SUSY} V^A_\mu &=-i\bar\xi \Gamma_\mu \lambda^A, \label{eq:4dSUSYtransfA}\\
		\delta_{\rm SUSY} V^A_p &=-i\bar\xi \Gamma_p \lambda^A, \label{eq:4dSUSYtransfB}\\
		\delta_{\rm SUSY} \lambda^A &=\left( \frac12 F^A_{\mu\nu} \Gamma^{\mu\nu} +F^A_{\mu p} \Gamma^{\mu p} +\frac12 F^A_{pq} \Gamma^{pq} \right)\xi \label{eq:4dSUSYtransfC}.
	\end{align}
\end{subequations}

In terms of the 4D fields from \eqref{eq:KKspinor} and \eqref{eq:KKvector} and the 4D spinor parameters from \eqref{eq:xiSUSYparam}, the transformation \eqref{eq:4dSUSYtransfA} gives
\begin{equation}\label{eq:SUSYtransfAvec}
	\begin{split}
		\delta_{\rm SUSY} A^{A0}_\mu &=-\frac{i}{\sqrt{V}} \sum_\alpha \Big( \bar\xi_{L,\alpha} \gamma_\mu \psi^{A\alpha}_{L} +\bar\xi_{R,\alpha} \gamma_\mu \psi^{A\alpha}_{R} \Big), \\
		\delta_{\rm SUSY} A^{Aa}_\mu &=-i\sum_{\alpha,i} \Big( c_{\alpha i a} \, \bar\xi^\alpha_{L} \gamma_\mu \psi^{Ai}_{L} +c_{i\alpha a} \, \bar\xi^\alpha_{R} \gamma_\mu \psi^{Ai}_{R} \Big).
	\end{split}
\end{equation}
Using \eqref{eq:Lich} and \eqref{eq:6dRicciFlat} we find that
\begin{equation}
	\big( \lambda_a -\lambda_i^2 \big)c_{\alpha i a} =0,
\end{equation}
where there is no sum over $i$ and $a$. This implies that the triple overlaps vanish when the 4D massive vector and the 4D massive spinor have different masses. The SUSY transformation of a massive vector thus only involves fermions with the same mass as the vector. This is expected, since fields belonging to the same supersymmetric multiplet have equal mass in flat spacetime.

Similarly, \eqref{eq:4dSUSYtransfB} yields
\begin{align}
	\delta_{\rm SUSY} A^{Aa} &=		\frac1{\lambda_a}\sum_{\alpha,i} \lambda_i\Big( c_{\alpha i a} \, \bar\xi^\alpha_{L} \psi^{Ai}_{R} -c_{i\alpha a} \, \bar\xi^\alpha_{R} \psi^{Ai}_{L} \Big), \label{eq:SUSYtransfAscalar1} \\
	\delta_{\rm SUSY} \phi^{AI} &=-i\sum_{\alpha,i} \Big( c_{\alpha iI} \, \bar\xi^\alpha_{L} \psi^{Ai}_{R} +\big(c_{\alpha iI}\big)^* \, \bar\xi^\alpha_{R} \psi^{Ai}_{L} \Big).\label{eq:SUSYtransfAscalar2}
\end{align}
 As before, we find that
\begin{equation}
	\big( \lambda_I -\lambda_i^2 \big)c_{\alpha i I} =0,
\end{equation}
where there is no sum over $i$ and $I$.

Let us split the three terms of \eqref{eq:4dSUSYtransfC} into
\begin{equation}
	\delta_{\rm SUSY} \lambda^A =\delta^{(1)}_{\rm SUSY} \lambda^A +\delta^{(2)}_{\rm SUSY} \lambda^A +\delta^{(3)}_{\rm SUSY} \lambda^A.
\end{equation}

In $d=4$, we have
\begin{equation}
	\delta_{\rm SUSY} \psi^A_{L,i} =\delta_{\rm SUSY} ^{(1)}\psi^A_{L,i} +\delta_{\rm SUSY} ^{(2)}\psi^A_{L,i} +\delta_{\rm SUSY} ^{(3)}\psi^A_{L,i}.
\end{equation}

The first two terms read
\begin{align}
	\delta_{\rm SUSY} ^{(1)}\psi^A_{L,i} &=\frac12 \sum_\alpha \bigg[ \frac{\delta_{i\alpha}}{\sqrt{V}} X^{A0}_{\mu\nu} +\sum_a c_{i\alpha a}X^{Aa}_{\mu\nu} \bigg]\gamma^{\mu\nu}\xi^\alpha_{L}, \label{eq:SUSYspinor1} \\
	\delta_{\rm SUSY} ^{(2)}\psi^A_{L,i} &=\sum_\alpha \bigg[ -i\lambda_i\sum_a c_{\alpha i a} X^{Aa}_\mu +\sum_I c_{i\alpha I} X^{AI}_\mu \bigg]\gamma^\mu\xi^\alpha_{R}, \label{eq:SUSYspinor2}
\end{align}
with
\begin{equation}
	\begin{split}
		X^{A0}_{\mu\nu} &\coloneq\partial_\mu A^{A0}_\nu -\partial_\nu A^{A0}_\mu +\frac{g}{\sqrt{V}}{f^A}_{BC}A^{B0}_\mu A^{C0}_\nu +\frac{g}{\sqrt{V}}{f^A}_{BC}\sum_a A^{Ba}_\mu A^{Ca}_\nu, \\
		X^{Aa}_{\mu\nu} &\coloneq\partial_\mu A^{Aa}_\nu -\partial_\nu A^{Aa}_\mu +\frac{g}{\sqrt{V}}{f^A}_{BC}\big(A^{Ba}_\mu A^{C0}_\nu +A^{B0}_\mu A^{Ca}_\nu \big) +g{f^A}_{BC}\sum_{b,c} c_{abc} A^{Bb}_\mu A^{Cc}_\nu,
	\end{split}
\end{equation}
and
\begin{equation}
	\begin{split}
		X^{Aa}_{\mu} &\coloneq\partial_\mu A^{Aa} -A^{Aa}_\mu
		+ \frac{g}{\sqrt{V}} {f^A}_{BC} A^{B0}_\mu A^{Ca} +g{f^A}_{BC} \sum_{b,c} \frac{\lambda_a +\lambda_c -\lambda_b}{2\lambda_a} c_{abc}\, A^{Bb}_\mu A^{Cc} \\
		&\quad +g{f^A}_{BC} \sum_{b,J} \frac{c_{abJ}}{\lambda_a}\, A^{Bb}_\mu \phi^{CJ}, \\
		X^{AI}_{\mu} &\coloneq\partial_\mu \phi^{AI}
		+\frac{g}{\sqrt{V}}{f^A}_{BC} A^{B0}_\mu \phi^{CI} +g{f^A}_{BC} \sum_{b,c} c_{cbI}\, A^{Bb}_\mu A^{Cc}
		+g{f^A}_{BC} \sum_{b,J} c_{bIJ}\, A^{Bb}_\mu \phi^{CJ}.
	\end{split}
\end{equation}

From the last term of \eqref{eq:4dSUSYtransfC}, we get the transformation
\begin{equation}\label{eq:SUSYspinor3}
	\begin{split}
		\delta_{\rm SUSY} ^{(3)}\psi^A_{L,i} &=\sum_\alpha \bigg[ -i\lambda_i \sum_I \big( c_{\alpha iI} \big)^* \bigg( \phi^{AI} -\frac12g{f^A}_{BC}\sum_{b,c} c_{bcI} A^{Bb}A^{Cc} \bigg) \\
		&\quad\quad\quad\quad +ig{f^A}_{BC} \sum_{a,I} \bigg( \sum_J \lambda_i  c_{aIJ}\big( c_{i\alpha J} \big)^* -\sum_k\lambda_k\,c_{ika}\big( c_{k\alpha I} \big)^* \bigg) A^{Ba}\phi^{CI} \\
		&\quad\quad\quad\quad\quad +\frac12g{f^A}_{BC} \sum_{I,J,k} c_{ikI}\big( c_{\alpha kJ} \big)^* \, \phi^{BI}\phi^{CJ} \bigg]\xi^\alpha_{L}.
	\end{split}
\end{equation}
The SUSY transformations of the 4D spinors are given by the sum of \eqref{eq:SUSYspinor1}, \eqref{eq:SUSYspinor2} and \eqref{eq:SUSYspinor3}.

Lastly, let us comment on the SUSY multiplets of the KK modes when the internal manifold is a torus or a Calabi--Yau manifold.
When $\mathcal{N}_4=4$, the 4D spectrum consists of, for each adjoint index, a massless vector multiplet plus a $\frac12$-BPS $\mathcal{N}_4=4$ massive vector multiplet for each nonzero momentum $\vec{k}\in\Lambda^*$, where the torus is $\mathbb{R}^6/\Lambda$.
When $\mathcal{N}_4=1$, the 4D theory has an $\mathcal{N}_4=1$ massless vector multiplet, an $\mathcal{N}_4=1$ massive vector multiplet for each non-constant scalar Laplace eigenmode, and an $\mathcal{N}_4=1$ massive chiral multiplet for each co-exact $(0,1)$-eigenform on the Calabi--Yau threefold.

\section{Scattering amplitudes in four dimensions}\label{sec:3}

We now turn to computing scattering amplitudes in the dimensionally reduced theory. In particular, we compute tree-level $2 \rightarrow 2$ amplitudes that include fermionic KK modes as external states. If these amplitudes grow like $E^n$ in the high-energy limit with $n>0$, we obtain sum rules by requiring that the growing parts vanish, as is needed to match the $E^0$ high-energy behavior of the 10D theory. These sum rules give instances of the unitarity sum rules for spontaneously broken gauge theories with fermions \cite{LlewellynSmith:1973yud, Gunion:1990kf}.

\subsection{Kinematics}

We consider $2 \rightarrow 2$ scattering in the center-of-mass frame. We take the external momentum of the $i$th particle, $i=1,2,3,4$,  to be
\begin{equation}
	p^\mu_i(p_i)=(E_i,p_i\sin\theta_i,0,p_i\cos\theta_i),
\end{equation}
where $p_i$ denotes the norm of the 3-momentum and $\theta_i$ the scattering angle. We also have $E^2_i=p_i^2+m^2_i$ and
\begin{equation}
	\theta_1=0, \quad \theta_2=\pi, \quad \theta_3=\theta, \quad \theta_4=\theta-\pi.
\end{equation}
Momentum conservation gives $E_1+E_2=E_3+E_4=E$, $p_1=p_2$, and $p_3=p_4$.

The longitudinal and transverse vector polarizations are
\begin{equation}
	\begin{split}
		\epsilon_L^\mu(p_i) &=\frac1{m_i}(p_i,E_i\sin\theta_i,0,E_i\cos\theta_i), \\
		\epsilon_\pm^\mu(p_i) &=\frac1{\sqrt{2}}(0,\mp\cos\theta_i,-i,\pm\sin\theta_i),
	\end{split}
\end{equation}
where $\epsilon_\pm^\mu(p_i)$ are definite helicity polarizations.

Following Srednicki's book \cite{Srednicki:2007qs}, the massive spinor polarizations read
\begin{equation}
	u_{\uparrow,\downarrow}(p_i) =\frac1{\sqrt{2(E_i+m_i)}}(\slashed{p_i}-im_i)u_{\uparrow,\downarrow}(0),
\end{equation}
where $\slashed{p_i}\coloneq p^\mu_i\gamma_\mu$,
\begin{equation}
	\begin{split}
		u_{\uparrow}(0) &=\cos\alpha\,u_1(0)+\sin\alpha\,u_2(0), \\
		u_{\downarrow}(0) &=-\sin\alpha\,u_1(0)+\cos\alpha\,u_2(0),
	\end{split}
\end{equation}
for $\alpha\coloneq\frac{\theta_i}2-\frac\pi4$ and
\begin{equation}
	u_1(0)=\begin{pmatrix}
		1\\0\\-i\\0
	\end{pmatrix}, \quad u_2(0)=\begin{pmatrix}
		0\\1\\0\\i
	\end{pmatrix}.
\end{equation}
The Majorana condition then fixes
\begin{equation}
	v^*_s(p_i) =B_4u_s(p_i).
\end{equation}

\subsection{Amplitudes}
We consider $2 \rightarrow 2$ processes with at least two external fermion particles, since otherwise the tree-level amplitudes are purely bosonic and reduce to those considered in \cite{Bonifacio:2019ioc}. 
The relevant off-shell  cubic vertices involving fermions are\footnote{In this section only, we use fraktur letters $\mathfrak{a},\mathfrak{b}$ to denote spinor indices in $d=4$, which we usually suppress.}
\begin{equation}
	\begin{split}
		\mathcal{V} \big(A^{Ba}_\mu,\psi^{C}_{i\mathfrak{b}},\psi^{A}_{j\mathfrak{a}}\big) &=gf^{ABC} \Big[B_4^T\gamma^0\gamma_\mu\big( c_{jia}P_L +(c_{jia})^*P_R \big)\Big]_{\mathfrak{a}\mathfrak{b}}, \\
		\mathcal{V} \big(A^{B0}_\mu,\psi^{C}_{i\mathfrak{b}},\psi^{A}_{j\mathfrak{a}}\big) &=\frac{g}{\sqrt{V}}f^{ABC}\,\delta_{ij} \Big[B_4^T\gamma^0\gamma_\mu\Big]_{\mathfrak{a}\mathfrak{b}}, \\
		\mathcal{V} \big(\phi^{BI},\psi^{C}_{i\mathfrak{b}},\psi^{A}_{j\mathfrak{a}}\big) &=igf^{ABC} \Big[B_4^T\gamma^0\big( -c_{jiI}P_R +(c_{jiI})^*P_L \big)\Big]_{\mathfrak{a}\mathfrak{b}},
	\end{split}
\end{equation}
where all momenta are incoming.
The purely bosonic vertices, involving only the massive vectors $A^{Aa}_\mu$, the massless vectors $A^{A0}_\mu$ and the scalars $\phi^{AI}$, are given in \cite{Bonifacio:2019ioc}.

\subsubsection{Vector-spinor scattering}

Let us first consider the $2\to2$ scattering amplitude of a spinor with a massive vector,
\begin{equation}\label{eq:process}
	\psi^A_j(p_1,s_1) A^{Ba}_{r_2}(p_2)\to \psi^C_i(p_3,s_3) A^{Db}_{r_4}(p_4),
\end{equation}
where the labels $s_i=\,\,\uparrow,\downarrow$ indicate the spinor polarizations and $r_i=\pm,L$ denote the transverse or longitudinal vector polarizations.

All three channels, $s$, $t$, and $u$, contribute to this process. The total amplitude is
\begin{equation} \label{eq:TotalAmp}
	\mathcal{A} =\mathcal{A}_s +\mathcal{A}_u +\mathcal{A}^{(1)}_t +\mathcal{A}^{(2)}_t +\mathcal{A}^{(3)}_t,
\end{equation}
where the $s$- and $u$-channel contributions come from exchanged KK fermions and the three $t$-channel contributions come from the exchange of massive KK vectors, massless vectors, and KK scalars, respectively. Using spring-like lines to represent massive vectors, the $s$ and $u$ channel Feynman diagrams take the following form:
\begin{center}
	\begin{tikzpicture}
		\begin{feynman}
			\vertex (i) at (-3,0) {\(A,j\)};
			\vertex (v1) at (-1,0);
			\vertex (v2) at (1,0);
			\vertex (f) at (3,0) {\(C,i\)};
			\vertex (g1) at (-1,-2) {\(B,a\)};
			\vertex (g2) at (1,-2) {\(D,b\)};
			\diagram* {
				(i) -- [fermion, momentum={$p_1$}] (v1)
				-- [fermion, edge label'={\(F,k\)}, momentum={$p_1+p_2$}] (v2)
				-- [fermion, momentum={$p_3$}] (f),
				(g1) -- [gluon, momentum={$p_2$}] (v1),
				(v2) -- [gluon, momentum={$p_4$}] (g2),
			};
		\end{feynman}
	\end{tikzpicture} \quad\quad
	\begin{tikzpicture}
		\begin{feynman}
			\vertex (i) at (-3,0) {\(A,j\)};
			\vertex (v1) at (-1,0);
			\vertex (v2) at (1,0);
			\vertex (f) at (3,0) {\(C,i\)};
			\vertex (g1) at (-1,-2) {\(D,b\)};
			\vertex (g2) at (1,-2) {\(B,a\)};
			\diagram* {
				(i) -- [fermion, momentum={$p_1$}] (v1)
				-- [fermion, edge label'={\(F,k\)}, momentum={$p_1-p_4$}] (v2)
				-- [fermion, momentum={$p_3$}] (f),
				(v1) -- [gluon, momentum'={$p_4$}] (g1),
				(g2) -- [gluon, momentum'={$p_2$}] (v2),
			};
		\end{feynman}
	\end{tikzpicture}
\end{center}
Using wavy lines to represent massless vectors and dashed lines to represent scalars, the three types of $t$ channel diagram take the following form:
\begin{center}
	\begin{tikzpicture}[scale=0.7]
		\begin{feynman}
			\vertex (a1) at (-3.0,  1.0);
			\vertex (a2) at ( 3.0,  1.0);
			\vertex (b1) at (-3.0,  0.0) {\(A,j\)};
			\vertex (v1) at ( 0.0,  0.0);
			\vertex (b2) at ( 3.0,  0.0) {\(C,i\)};
			\vertex (v2) at ( 0.0, -2.6);
			\vertex (g1) at (-2.6, -5.2) {\(B,a\)};
			\vertex (g2) at ( 2.6, -5.2) {\(D,b\)};
			
			\diagram* {
				(b1) -- [fermion, momentum={$p_1$}]  (v1)
				-- [fermion, momentum={$p_3$}] (b2),			
				(v1) -- [gluon, edge label'={\(F,c\,\,\)}, momentum={$p_1-p_3$}]    (v2),
				(g1) -- [gluon, momentum={$p_2$}] (v2),
				(v2) -- [gluon, momentum={$p_4$}]  (g2),
			};
		\end{feynman}
	\end{tikzpicture}\quad\,\,
	\begin{tikzpicture}[scale=0.7]
		\begin{feynman}
			\vertex (a1) at (-3.0,  1.0);
			\vertex (a2) at ( 3.0,  1.0);
			\vertex (b1) at (-3.0,  0.0) {\(A,j\)};
			\vertex (v1) at ( 0.0,  0.0);
			\vertex (b2) at ( 3.0,  0.0) {\(C,i\)};
			\vertex (v2) at ( 0.0, -2.6);
			\vertex (g1) at (-2.6, -5.2) {\(B,a\)};
			\vertex (g2) at ( 2.6, -5.2) {\(D,b\)};
			
			\diagram* {
				(b1) -- [fermion, momentum={$p_1$}]  (v1)
				-- [fermion, momentum={$p_3$}] (b2),			
				(v1) -- [photon, edge label'={\(F\,\)}, momentum={$p_1-p_3$}]    (v2),
				(g1) -- [gluon, momentum={$p_2$}] (v2),
				(v2) -- [gluon, momentum={$p_4$}]  (g2),
			};
		\end{feynman}
	\end{tikzpicture}\quad\,\,
	\begin{tikzpicture}[scale=0.7]
		\begin{feynman}
			\vertex (a1) at (-3.0,  1.0);
			\vertex (a2) at ( 3.0,  1.0);
			\vertex (b1) at (-3.0,  0.0) {\(A,j\)};
			\vertex (v1) at ( 0.0,  0.0);
			\vertex (b2) at ( 3.0,  0.0) {\(C,i\)};
			\vertex (v2) at ( 0.0, -2.6);
			\vertex (g1) at (-2.6, -5.2) {\(B,a\)};
			\vertex (g2) at ( 2.6, -5.2) {\(D,b\)};
			
			\diagram* {
				(b1) -- [fermion, momentum={$p_1$}]  (v1)
				-- [fermion, momentum={$p_3$}] (b2),			
				(v1) -- [scalar, edge label'={\(F,I\,\)}, momentum={$p_1-p_3$}]    (v2),
				(g1) -- [gluon, momentum={$p_2$}] (v2),
				(v2) -- [gluon, momentum={$p_4$}]  (g2),
			};
		\end{feynman}
	\end{tikzpicture}
\end{center}

The explicit expressions for the contributions in \eqref{eq:TotalAmp} are given by
\footnotesize
\begin{align}
		i\mathcal{A}_s &=g^2 \sum_{k} f_{CDF}{f^F}_{BA} \bar{u}_{s_3}(p_3) \Big[ \gamma^\mu\big( c_{ikb}P_L+(c_{ikb})^*P_R \big) \Big]  \frac{-i[-(\slashed{p}_1+\slashed{p}_2)+i\lambda_k]}{(p_1+p_2)^2+\lambda_k^2} \Big[ \gamma^\nu\big( c_{kja}P_L+(c_{kja})^*P_R \big) \Big]u_{s_1}(p_1) \nonumber \\
		&\quad \epsilon_{r_2\nu}(p_2) \epsilon^*_{r_4\mu}(p_4), \\
		i\mathcal{A}_u &=g^2 \sum_{k} f_{CBF}{f^F}_{DA} \bar{u}_{s_3}(p_3) \Big[ \gamma^\mu\big( c_{ika}P_L+(c_{ika})^*P_R \big) \Big]  \frac{-i[-(\slashed{p}_1-\slashed{p}_4)+i\lambda_k]}{(p_1-p_4)^2+\lambda_k^2} \Big[ \gamma^\nu\big( c_{kjb}P_L+(c_{kjb})^*P_R \big) \Big]u_{s_1}(p_1) \nonumber \\
		&\quad \epsilon_{r_2\mu}(p_2) \epsilon^*_{r_4\nu}(p_4), \\
		i\mathcal{A}^{(1)}_t &=-g^2\sum_{c}f_{CFA}{f^F}_{BD}\,c_{abc}\,\bar{u}_{s_3}(p_3) \Big[ \gamma^\mu\left( c_{ijc}P_L+(c_{ijc})^*P_R \right) \Big] u_{s_1}(p_1) \frac{-i\left(\eta_{\mu\nu}+\frac{(p_1-p_3)_\mu (p_1-p_3)_\nu}{\lambda_c}\right)}{(p_1-p_3)^2+\lambda_c} \nonumber \\
		&\quad \Big( \eta^{\nu\rho}(p_4-2p_2)^\sigma +\eta^{\rho\sigma}(p_2+p_4)^\nu +\eta^{\sigma\nu}(p_2-2p_4)^\rho \Big) \epsilon_{r_2\rho}(p_2)\epsilon^*_{r_4\sigma}(p_4), \\
		i\mathcal{A}^{(2)}_t &=-\frac{g^2}{V}\delta_{ij}\delta_{ab} f_{CFA}{f^F}_{BD}\bar{u}_{s_3}(p_3) \gamma^\mu u_{s_1}(p_1) \frac{-i\eta_{\mu\nu}}{(p_1-p_3)^2} \nonumber \\
		&\quad \Big( \eta^{\nu\rho}(p_4-2p_2)^\sigma +\eta^{\rho\sigma}(p_2+p_4)^\nu +\eta^{\sigma\nu}(p_2-2p_4)^\rho \Big)\epsilon_{r_2\rho}(p_2)\epsilon^*_{r_4\sigma}(p_4), \\
		i\mathcal{A}^{(3)}_t &=-2g^2\sum_{I}f_{CFA}{f^F}_{BD}\,c_{abI}\,\bar{u}_{s_3}(p_3) \Big[\left(-c_{ijI}P_R +(c_{ijI})^*P_L \right) \Big] u_{s_1}(p_1) \frac{-i}{(p_1-p_3)^2+\lambda_I} \epsilon^{\mu}_{r_2}(p_2)\epsilon^*_{r_4\mu}(p_4).
\end{align}
\normalsize

Let us first consider the case where the two external massive vectors are longitudinally polarized, i.e., $r_2=r_4=L$. From power counting, we see that for generic values of the masses and the coupling constants, the high-energy behavior of the amplitude $\mathcal{A}$ is $\sim E^2$, where $E=E_1+E_2$ is the center-of-mass energy. Since $2 \rightarrow 2$ tree-level amplitudes in $D=10$ SYM grow at most like $\sim E^0$ at high energies, the overall $E^2$ and $E^1$ contributions to  $\mathcal{A}$ must cancel due to nontrivial cancellations between Feynman diagrams.  

Setting to zero the leading $E^2$ contribution to the amplitude with $s_1=s_3=\,\,\uparrow$, and making use of the Jacobi identity
\begin{equation}\label{eq:Jacobi}
	f_{ABF}{f_{CD}}^F -f_{ACF}{f_{BD}}^F +f_{ADF}{f_{BC}}^F =0,
\end{equation}
we get
\begin{equation}\label{eq:SR1Pre}
	\begin{split}
		0 &=\frac{g^2E^2}{\sqrt{\lambda_a\lambda_b}}\cos\frac\theta2\bigg[f_{CBF}{f_{DA}}^{F}\sum_k\big(c_{jka}\,c_{kib}-c_{jkb}\,c_{kia} \big) \\
		&\quad +f_{CFA}{f_{BD}}^{F}\bigg(\frac1V\delta_{ab}\delta_{ij} +\sum_c c_{abc}\,c_{jic} -\sum_k c_{jka}\, c_{kib} \bigg) \bigg].
	\end{split}
\end{equation}
The second line of \eqref{eq:SR1Pre} gives the following complex sum rule:
\begin{equation}\label{eq:SRApsiApsi}
	0 =\frac1V\delta_{ab}\delta_{ij} +\sum_c c_{abc}\,c_{jic} -\sum_k c_{jka}\, c_{kib}.
\end{equation}
The sums over $c$ and $k$ correspond to summing over the exchanged spin-$1$ and spin-$\frac{1}{2}$ particles, respectively. There are no sums over vector eigenmodes since the scalar exchange diagrams do not contribute at order $E^2$. 

The first line of \eqref{eq:SR1Pre} vanishes as a consequence of \eqref{eq:SRApsiApsi}, since taking the antisymmetric part under $a\leftrightarrow b$ gives
\begin{equation}\label{eq:trivialSR}
	0 =\sum_k\big(c_{jkb}\,c_{kia} -c_{jka}\,c_{kib} \big).
\end{equation}

Considering the amplitude with $s_1=s_3=\,\,\downarrow$ gives the complex conjugate of \eqref{eq:SRApsiApsi}, and no new sum rules are obtained. The amplitudes with $s_1\ne s_3$ vanish at order $E^2$, and so no other sum rules are present at this order.

Let us now focus on contributions at order $E$. This time, there is no contribution when $s_1=s_3$, since this amplitude is invariant under $E \rightarrow -E$. Let us first consider the amplitude with $s_1=\,\,\uparrow$ and $s_3=\,\,\downarrow$. Setting to zero the $E^1$ contributions to the amplitude and using the Jacobi identity gives
\begin{equation}
	\begin{split}
		0 &=\frac{g^2E}{\sqrt{\lambda_a\lambda_b}}\sin\frac\theta2 \bigg[-f_{CBF}{f_{DA}}^{F}\sum_k \Big( c_{kia}c_{jkb}\, \lambda_i +c_{kjb}\, c_{ika}\, \lambda_j -\lambda_k\big(c_{ika}\,c_{jkb} +c_{ikb}\,c_{jka} \big)\Big) \\
		&\quad +f_{CFA}{f_{BD}}^{F} \bigg( \frac{(\lambda_i +\lambda_j)\delta_{ab}\delta_{ij}}{2V} +\sum_c \frac{c_{abc}\big(c_{jic}(\lambda_a-\lambda_b+\lambda_c)\lambda_i +c_{ijc}(-\lambda_a+\lambda_b+\lambda_c)\lambda_j\big)}{2\lambda_c} \\
		&\quad\quad\quad\quad\quad\quad\quad\quad -\sum_k \lambda_k\, c_{ikb}\, c_{jka} +i\sum_I c_{abI}\,c_{ijI} \bigg)\bigg].
	\end{split}
\end{equation}
This leads to the following two additional sum rules:
\begin{align}
	0 &=\sum_{k} \Big( c_{kia}\,c_{jkb}\, \lambda_i +c_{kjb}\, c_{ika}\, \lambda_j -\lambda_k\big(c_{ika}\,c_{jkb} +c_{ikb}\,c_{jka} \big)\Big), \label{eq:SRApsiApsiE1a} \\
	0 &=\sum_c \frac{\lambda_a-\lambda_b}{\lambda_c} c_{abc} \big( c_{jic}\,\lambda_i -c_{ijc}\,\lambda_j \big) +2i\sum_I c_{abI} c_{ijI} -\sum_k \lambda_k \big( c_{ikb}\,c_{jka} -c_{jkb}\,c_{ika} \big),\label{eq:SRApsiApsiE1b}
\end{align}
where we used \eqref{eq:SRApsiApsiE1a} and the previous sum rule \eqref{eq:SRApsiApsi} to simplify \eqref{eq:SRApsiApsiE1b}. Considering instead the amplitude with $s_1=\,\,\downarrow$ and $s_3=\,\,\uparrow$ leads to sum rules that are the complex conjugates of \eqref{eq:SRApsiApsiE1a} and \eqref{eq:SRApsiApsiE1b}.

After enforcing the sum rules, the leading non-vanishing part of the elastic amplitude for $s_1=s_3=\,\uparrow$ and $r_2=r_4=L$ is given explicitly by
\begin{equation}
	\begin{split}
		i\mathcal{A} &=-i \frac{g^2}{2\lambda_a \cos\tfrac{\theta}{2}} f_{ABF}{f^F}_{BA}  \sum_{k} \bigg[ c_{ika}\,c_{kia}\,\lambda_i^2 +\big(c_{ika}^2 +c_{kia}^2\big)\lambda_i\lambda_k -3\,c_{ika}\,c_{kia}\,\lambda_k^2 \\
		&\quad -\Big(c_{ika}\,c_{kia}\,\lambda_k^2 +\big(c_{ika}^2 +c_{kia}^2\big)\lambda_i\lambda_k -3\, c_{ika}\,c_{kia}\,\lambda_i^2\Big)\cos\theta \bigg] +O\big(E^{-1}\big).
	\end{split}
\end{equation}

If we consider amplitudes with one longitudinally polarized external massive vector and one transversely polarized external massive vector, e.g., $r_2=L,r_4=\pm$, then \eqref{eq:SRApsiApsi} is enough to unitarize the amplitude at leading order $E$ and so we do not obtain any new sum rules.

\subsubsection{Other processes}

Let us now consider other processes. From power counting, and assuming at least two external spinors, we need to have at least one longitudinally polarized external massive vector to achieve a high-energy growth of at least $E^1$. Only two processes might contribute, in addition to the vector-spinor scattering considered above. In both of these additional cases, the leading contribution is at order $E$:
\begin{enumerate}
\item Consider first the amplitude with an incoming fermion and massive longitudinally polarized vector and an outgoing spinor and massless vector,
\begin{equation}
	\psi^A_j(p_1,s_1) A^{Ba}_{L}(p_2)\to \psi^C_i(p_3,s_3) A^{D0}_{r_4}(p_4),
\end{equation}
where $r_4=\pm$. This amplitude vanishes at order $E$ by the Jacobi identity, so we do not obtain any new sum rule.
\item Consider now the amplitude with an outgoing scalar:
\begin{equation}
	\psi^A_j(p_1,s_1) A^{Ba}_{L}(p_2)\to \psi^C_i(p_3,s_3) \phi^{DI}(p_4).
\end{equation}
There is no contribution for $s_1=s_3$. For $s_1=\,\,\uparrow,s_3=\,\,\downarrow$, and after using \eqref{eq:Jacobi}, we get the following condition at order $E$:
\begin{equation}
	\begin{split} \label{eq:scaler_amp}
		0 &=\frac{g^2E}{\sqrt{\lambda_a}}\sin\frac\theta2 \bigg[f_{CBF}{f_{DA}}^{F}\sum_k\big(c_{ika}\,c_{kjI} +c_{jka}\,c_{kiI} \big) \\
		&\quad +f_{CFA}{f_{BD}}^{F}\bigg( i\sum_c \frac{c_{caI}\big( c_{jic}\,\lambda_i -c_{ijc}\,\lambda_j\big)}{\lambda_c} +\sum_J c_{aIJ}\,c_{jiJ} -\sum_k c_{jka}\,c_{kiI} \bigg) \bigg].
	\end{split}
\end{equation}
From the second line, we obtain the following additional sum rule:
\begin{equation}\label{eq:SRscalarExt}
	0 =i\sum_c \frac{c_{caI}\big( c_{jic}\,\lambda_i -c_{ijc}\,\lambda_j\big)}{\lambda_c} +\sum_J c_{aIJ}\,c_{jiJ} -\sum_k c_{jka}\,c_{kiI}.
\end{equation}
The first line of \eqref{eq:scaler_amp} then vanishes by taking the symmetric part under $i \leftrightarrow j$ of \eqref{eq:SRscalarExt} and using the symmetry properties of the triple overlaps. \end{enumerate}

There are no additional processes that naively grow at high energies. In particular, we do not get any sum rule from the spinor-spinor scattering process,
\begin{equation}
	\psi^A_j(p_1,s_1) \psi^B_k(p_2,s_2)\to \psi^C_i(p_3,s_3) \psi^D_l(p_4,s_4).
\end{equation}
This receives contributions from $s$, $t$ and $u$ channel diagrams involving the exchange of vectors and scalars. The leading $E^0$ part of the elastic amplitude, $i=j$ and $k=l$, with $s_1=s_2=s_3=s_4=\,\uparrow$ is given explicitly by
\begin{equation}
	\begin{split}
		i\mathcal{A} &=-i g^2 f_{ABF}{f^F}_{BA} \Bigg[ \frac4{1+\cos\theta}\frac{\delta_{jk}}V +\sum_I |c_{jkI}|^2  \\
		&\quad\quad\quad +\sum_c \bigg( \frac4{1+\cos\theta}\,|c_{jkc}|^2 +\frac{|c_{kjc}\,\lambda_j-c_{jkc}\lambda_k|^2}{\lambda_c} \bigg) \Bigg] +O\big(E^{-1}\big).
	\end{split}
\end{equation}

\subsection{Proving the sum rules}

In the previous section, we obtained the following four sum rules by imposing perturbative unitarity of scattering amplitudes:
\begin{subequations} \label{eq:allSumRules}
\begin{align}
	0 & =\frac1V\delta_{ab}\delta_{ij} +\sum_c c_{abc}\,c_{jic} -\sum_k c_{jka}\, c_{kib}, \label{eq:SRApsiApsi_2} \\
	0 &=\sum_{k} \Big( c_{kia}\,c_{jkb}\, \lambda_i +c_{kjb}\, c_{ika}\, \lambda_j -\lambda_k\big(c_{ika}\,c_{jkb} +c_{ikb}\,c_{jka} \big)\Big), \label{eq:SRApsiApsiE1a_2} \\
	0 &=\sum_c \frac{\lambda_a-\lambda_b}{\lambda_c} c_{abc} \big( c_{jic}\,\lambda_i -c_{ijc}\,\lambda_j \big) +2i\sum_I c_{abI} c_{ijI} -\sum_k \lambda_k \big( c_{ikb}\,c_{jka} -c_{jkb}\,c_{ika} \big),\label{eq:SRApsiApsiE1b_2} \\
	0 & =i\sum_c \frac{c_{caI}\big( c_{jic}\,\lambda_i -c_{ijc}\,\lambda_j\big)}{\lambda_c} +\sum_J c_{aIJ}\,c_{jiJ} -\sum_k c_{jka}\,c_{kiI}. \label{eq:SRscalarExt_2}
\end{align}
\end{subequations}
These are instances of the unitarity sum rules from \cite{Gunion:1990kf}, where here the masses and couplings have an interpretation in terms of the spectral data of a manifold.
The analogues of the first three of these sum rules appear in \cite{Schwinn:2004xa}, without the scalar sum in \eqref{eq:SRApsiApsiE1b_2}, for the case of a 5D theory compactified on an interval. In Appendix \ref{app:examples}, we verify a Dirac spinor version of \eqref{eq:SRApsiApsi_2} on flat tori and round spheres.

It is possible to derive the sum rules \eqref{eq:allSumRules} from a purely geometrical perspective, by using completeness and associativity to evaluate integrals of products of four eigenmodes on $\mathcal{M}$ in two different ways. Let us first state the following useful relations, which follow from completeness of the various eigenmodes:
\begin{subequations}
\label{eq:completeness}
\begin{align}
		\zeta_a \zeta_b &=\frac1V \delta_{ab} +\sum_c c_{abc}\,\zeta_c, \\
		\nabla_p\zeta_a \, \zeta_b &=\sum_c \frac{\lambda_a-\lambda_b+\lambda_c}{2\lambda_c} c_{abc}\,\nabla_p\zeta_c +\sum_I c_{abI}\,Y_{p,I}, \\
		\zeta_a Y_{p,I} &=\sum_c \frac{c_{caI}}{\lambda_c}\,\nabla_p \zeta_c +\sum_J c_{aIJ}\,Y_{p,J}, \\
		Y_{p,I}Y^p_J &=\frac1V \delta_{IJ} +\sum_c c_{cIJ}\,\zeta_c, \\
		\chi^\dagger_{+,j} \chi_{+,i} &=\frac1V \delta_{ij} +\sum_c c_{jic}\,\zeta_c, \\
		\zeta_a \chi_{+,i} &=\sum_j c_{jia}\,\chi_{+,j}, \\
		\chi^\dagger_{+,j} \tilde\Gamma_p \chi_{-,i} &=i \sum_c \frac{\lambda_i\,c_{jic} -\lambda_j\,c_{ijc}}{\lambda_c}\nabla_p\zeta_c +\sum_I c_{jiI} Y_{p,I}, \\
		Y_{p,I} \tilde\Gamma^p \chi_{-,i} &=\sum_j c_{jiI}\,\chi_{+,j},\\
		\nabla_p\zeta_a \tilde\Gamma^p \chi_{-,i} &=i\sum_j \big(\lambda_i\,c_{jia} -\lambda_j\,c_{ija}\big)\chi_{+,j},
\end{align}
\end{subequations}
where we recall that the sums over $j$ include harmonic spinors. 

To derive the sum rule \eqref{eq:SRApsiApsi_2}, we consider the integral over $\mathcal{M}$ of the quadruple product $ \zeta_a  \zeta_b  \chi^\dagger_{+,j}  \chi_{+,i}$. We derive an identity by using completeness to evaluate this integral in two different ways,
\begin{equation}
	\int_\mathcal{M} \wick{
		\c1 \zeta_a \c1 \zeta_b \c2 \chi^\dagger_{+,j} \c2 \chi_{+,i}
	} =\int_\mathcal{M} \wick{
		\c1 \zeta_a \c2 \zeta_b \c1 \chi^\dagger_{+,j} \c2 \chi_{+,i}
	},
\end{equation}
where the square brackets indicate the use of the completeness relations \eqref{eq:completeness}. By expanding the indicated pairs of eigenmodes and then using orthonormality to do the integral, we obtain precisely the sum rule \eqref{eq:SRApsiApsi_2}. 

Similarly, to derive the sum rule \eqref{eq:SRApsiApsiE1a_2}, we start from the identity
\begin{equation}
	\int_\mathcal{M} \wick{
		\c1 {\nabla_p\zeta_a} \c1 \zeta_b \c2 \chi^\dagger_{+,j} \c2{\tilde\Gamma^p\chi_{-,i}}
	} =\int_\mathcal{M} \wick{
		\c2 {\nabla_p\zeta_a} \c1 \zeta_b \c1 \chi^\dagger_{+,j} \c2{\tilde\Gamma^p\chi_{-,i}}
	},
\end{equation}
which gives
\begin{equation}\label{eq:workingSRa}
	\begin{split}
		0 &=\sum_c \frac{\lambda_a-\lambda_b+\lambda_c}{2\lambda_c} c_{abc} \big(\lambda_i\,c_{jic} -\lambda_j\,c_{ijc}\big) -i \sum_I c_{abI}\,c_{jiI}  -\sum_k c_{jkb}\big( \lambda_i\,c_{kia} -\lambda_k\,c_{ika} \big).
	\end{split}
\end{equation}
Note that the first two sums of \eqref{eq:workingSRa} are antisymmetric under $i\leftrightarrow j$. By taking the symmetric part of \eqref{eq:workingSRa} under $i\leftrightarrow j$, and subsequently making use of \eqref{eq:trivialSR}, we find the sum rule \eqref{eq:SRApsiApsiE1a_2}. Similarly, the sum rule \eqref{eq:SRApsiApsiE1b_2} can be obtained from the antisymmetric part of \eqref{eq:workingSRa} under $a\leftrightarrow b$ after using \eqref{eq:trivialSR}.

The sum rule involving a four-dimensional external scalar field \eqref{eq:SRscalarExt_2} is reproduced exactly from the identity
\begin{equation}
	\int_\mathcal{M} \wick{
		\c1 {\zeta_a} \c1 {Y_{p,I}} \c2 \chi^\dagger_{+,j} \c2{\tilde\Gamma^p \chi_{-,i}}
	} =\int_\mathcal{M} \wick{
		\c1 {\zeta_a} \c2 {Y_{p,I}} \c1 \chi^\dagger_{+,j} \c2{\tilde\Gamma^p \chi_{-,i}}
	}.
\end{equation}

\section{Conclusions}\label{sec:conclusions}

In this paper, we have considered the Kaluza--Klein reduction of 10D SYM on a product spacetime $\mathbb{R}^{1,3}\times\mathcal{M}$, where $\mathcal{M}$ is an arbitrary closed, orientable six-dimensional Riemannian spin manifold.  The resulting four-dimensional theory consists of infinite towers of interacting massive KK particles of spin $0$, $1/2$, and $1$.  The masses of these fields are determined by the eigenvalues of Laplace and Dirac operators on the internal space $\mathcal{M}$, and their cubic coupling constants are given by integrals of products of three eigenmodes. 

Perturbative unitarity dictates that $2 \rightarrow 2$ tree-level scattering amplitudes of the massive KK modes should be bounded in the high-energy fixed-angle limit, to match the high-energy behavior of the 10D theory. As a consequence, the masses and cubic couplings must satisfy the sum rules given in \eqref{eq:allSumRules}, which eliminate the parts of the amplitude that would otherwise violate this bound. These give instances of the general unitarity sum rules for spontaneously broken gauge theories \cite{LlewellynSmith:1973yud, Gunion:1990kf}. From a geometric perspective, the sum rules give consistency conditions that must be satisfied by the eigenspectra and eigenmode overlaps of Laplace and Dirac operators on closed spin manifolds in six dimensions, such as Calabi--Yau manifolds. As we showed, the sum rules also can be derived directly from spectral decompositions of quadruple integrals on $\mathcal{M}$, using associativity and completeness. We also checked the simplest sum rule using explicit expressions for the eigenvalues and eigenmodes on flat tori and round spheres.

It is difficult to explicitly calculate the eigenvalues and eigenmode overlaps for most manifolds. The exceptions are highly symmetric spaces, such as flat tori and the round sphere considered in Appendix \ref{app:examples}. For Calabi--Yau threefolds, numerical methods must be used \cite{Braun:2008jp,Ashmore:2020ujw}. Thus, in general, the sum rules involve infinite sums over products of unknown cubic coupling constants. Nevertheless, we can hope to extract nontrivial information by using the approach of the conformal bootstrap \cite{Belavin:1984vu,Rattazzi:2008pe,Poland:2018epd}, as applied to manifolds and orbifolds \cite{Bonifacio:2020xoc, Bonifacio:2021msa,Bonifacio:2021aqf, Kravchuk:2021akc, Radcliffe:2024jcg, Bonifacio:2023ban, Gesteau:2023brw, Adve:2025rvf, Bertucci:2026fjh}. It is also possible to use KK sum rules to find bounds for more general classes of compactifications \cite{Mirbabayi:2026saz, Bertucci:2026osd}.  The sum rules in \eqref{eq:allSumRules} are likely too weak by themselves to extract strong upper bounds, which is a reflection of the fact that they apply to a large class of manifolds. However, we can still deduce some interesting consequences. Taking $i=j$ and $a=b$ in \eqref{eq:SRApsiApsiE1a_2}, we get 
\be
0= \sum_{k} \Big(|c_{ika}|^2\, \lambda_i -\lambda_k\big(c_{ika}\big)^2\Big).
\ee
Taking the real part and using ${\rm Re}(z^2) \leq |z|^2$ with the convention that $\lambda_k \geq 0$, we get
\be \label{eq:inequality}
0\leq \sum_{k} \left(\lambda_k-\lambda_i\right) |c_{ika}|^2.
\ee
For a given $i$ and $a$, suppose that the fermion $\psi_i$ couples through the massive vector $A_{\mu}^a$ to a heavier fermion $\psi_{k'}$, i.e., that  $c_{ik'a} \neq 0$ with $\lambda_{k'} > \lambda_i$.\footnote{By \eqref{eq:inequality}, it is not possible for $\psi_i$ to only couple through $A_{\mu}^a$ to fermions that are lighter than or have the same mass as $\psi_i$, unless all of them have the same mass as $\psi_i$.} Then applying \eqref{eq:inequality} to this heavier fermion $\psi_{k'}$ shows that it must couple through $A_{\mu}^a$ to an even heavier fermion (since by assumption it couples to a lighter fermion), which in turn must couple through $A_{\mu}^a$ to an even heavier fermion, and so on ad infinitum. There is thus an infinite chain of fermions that couple to heavier fermions through $A_{\mu}^a$.

A natural future direction, which partly motivated the present work, is to consider tree-level amplitudes in general compactifications of supergravity theories, building on the general KK compactifications for free massless spin-3/2 particles recently presented by one of the authors in \cite{Bertucci:2026omk}.  This is expected to lead to stronger sum rules for Ricci-flat spin manifolds, involving the eigenvalues of the Rarita--Schwinger operator. It may then be possible to use these sum rules to obtain interesting bootstrap bounds, giving nontrivial constraints on the spectra of massive gravitini in string compactifications.

\paragraph{Acknowledgements:}

We would like to thank Kurt Hinterbichler for helpful discussions. 

\paragraph{AI usage statement:}
The writing and calculations in this paper were done by the authors. We used Claude Fable 5.2 to check our results and to find errors in the draft.

\appendix

\section{Clifford algebras in even dimensions}\label{app:clifford}

In this appendix, we discuss gamma matrices in various dimensions, following the conventions of \cite{Polchinski_1998}. Gamma matrices in arbitrary dimension $D$ are denoted by $\Gamma^M_D$. In the main text, 4D gamma matrices are denoted by $\gamma^\mu$, while 6D Euclidean matrices are denoted by $\tilde\Gamma^p$. The $2^{\lfloor D/2 \rfloor}$-dimensional identity matrix in $D$ dimensions is denoted by $I_D$.

The Clifford algebra for $D$-dimensional Minkowski spacetime is
\begin{equation}
	\{\Gamma^M_D,\Gamma^N_D\} = 2\eta^{MN}I_D,
\end{equation}
where we use the mostly plus signature for the metric.
In even dimensions $D=2k+2$, we can define the chirality matrix,
\begin{equation}
	\Gamma_D =i^{-k}\Gamma^0_D \Gamma^1_D\dots\Gamma^{D-1}_D,
\end{equation}
which satisfies
\begin{equation}
	\{\Gamma_D,\Gamma^M_D\} = 0.
\end{equation}
We also define the completely antisymmetrized products,
\begin{equation}
	\Gamma^{MN\cdots Q} \coloneq \Gamma^{[M}_D \Gamma^N_D \cdots \Gamma^{Q]}_D,
\end{equation}
where we antisymmetrize with weight 1.
The Hermitian conjugate of a gamma matrix is
\begin{equation}
	\left( \Gamma_D^{M} \right)^{\dagger} =-\Gamma^0_D\Gamma^M_D (\Gamma^0_D)^{-1},
\end{equation}
and the Dirac adjoint of a spinor $\psi$ is
\begin{equation}
	\bar\psi \coloneq \psi^\dagger\Gamma^0_D.
\end{equation}

For even $D$, the gamma matrices can be constructed iteratively as follows: starting with $D=2$, we have
\begin{equation}
	\Gamma^0_{2} =\begin{pmatrix}
		0 & 1 \\ -1 & 0
	\end{pmatrix}, \qquad \Gamma^1_{2} =\begin{pmatrix}
		0 & 1 \\ 1 & 0
	\end{pmatrix}.
\end{equation}
Then for $D=2k+2$, we define
\begin{equation}
	\begin{split}
		\Gamma^\mu_D &=\Gamma^\mu_{D-2} \otimes \begin{pmatrix}
			-1 & 0 \\ 0 & 1
		\end{pmatrix}, \qquad \mu=0,\dots,D-3, \\
		\Gamma^{D-2}_D &=I_{D-2} \otimes \begin{pmatrix}
			0 & 1 \\ 1 & 0
		\end{pmatrix}, \qquad \Gamma^{D-1}_D =I_{D-2} \otimes \begin{pmatrix}
			0 & -i \\ i & 0
		\end{pmatrix},
	\end{split}
\end{equation}
where $\otimes$ indicates the Kronecker product between matrices and $\Gamma^\mu_{D-2}$ are $2^k \times 2^k$ gamma matrices in $D-2$ dimensions. 

For $D=4$ ($k=1)$, this gives
\begin{equation}
	\begin{split}
		\Gamma^0_{4} &=\begin{pmatrix}
			0 & 0 & -1 & 0 \\
			0 & 0 & 0 & 1 \\
			1 & 0 & 0 & 0 \\
			0 & -1 & 0 & 0
		\end{pmatrix}, \qquad \Gamma^1_{4} =\begin{pmatrix}0 & 0 & -1 & 0 \\
			0 & 0 & 0 & 1 \\
			-1 & 0 & 0 & 0 \\
			0 & 1 & 0 & 0
		\end{pmatrix}, \\
		\Gamma^{2}_{4} &=\begin{pmatrix}
			0 & 1 & 0 & 0 \\
			1 & 0 & 0 & 0 \\
			0 & 0 & 0 & 1 \\
			0 & 0 & 1 & 0
		\end{pmatrix}, \qquad \Gamma^{3}_{4} =\begin{pmatrix}
			0 & -i & 0 & 0 \\
			i & 0 & 0 & 0 \\
			0 & 0 & 0 & -i \\
			0 & 0 & i & 0
		\end{pmatrix}.
	\end{split}
\end{equation}
The chirality matrix in 4D is 
\be
\Gamma_{4} =\begin{pmatrix}
			1 & 0 & 0 & 0 \\
			0 & -1 & 0 & 0 \\
			0 & 0 & -1 & 0 \\
			0 & 0 & 0 & 1
		\end{pmatrix}.
\ee

For $D=6$ ($k=2)$, we get
\begin{equation}\label{eq:6dGammaMat}
	\begin{split}
		\Gamma_{6}^\mu &=\Gamma_{4}^\mu \otimes \begin{pmatrix}
			-1 & 0 \\ 0 & 1
		\end{pmatrix}, \qquad \mu=0,\dots,3, \\
		\Gamma_{6}^{4} &=I_4 \otimes \begin{pmatrix}
			0 & 1 \\ 1 & 0
		\end{pmatrix}, \qquad \Gamma_{6}^{5} =I_4 \otimes \begin{pmatrix}
			0 & -i \\ i & 0
		\end{pmatrix},
	\end{split}
\end{equation}
and the chirality matrix is
\be
\Gamma_{6} =\Gamma_{2}\otimes\Gamma_{4},
\ee
where the chirality matrix in 2D is
\be
 \Gamma_{2} =\Gamma^0_{2}   \Gamma^1_{2}=\begin{pmatrix}
			1 & 0 \\
			0 & -1
		\end{pmatrix}.
		\ee
We can define 6D Euclidean gamma matrices as
\begin{equation}
	\begin{split}
		\tilde\Gamma^0_{6} &\coloneq -i\Gamma^0_{6}, \\
		\tilde\Gamma^p_{6} &\coloneq \Gamma^p_{6}, \qquad p=1,\dots,5.
	\end{split}
\end{equation}
These are Hermitian matrices that satisfy the Euclidean Clifford algebra in six dimensions:
\begin{equation}
	\{\tilde\Gamma^p_{6},\tilde\Gamma^q_{6}\} =2\delta^{pq}I_6.
\end{equation}
We can also define the Euclidean chirality matrix,
\begin{equation}
	\tilde\Gamma_{6} =-\Gamma_{6},
\end{equation}
which satisfies
\begin{equation}
	\{\tilde\Gamma^p_{6},\tilde\Gamma_{6}\} =0, \qquad p=0,\dots,5.
\end{equation}

For $D=10$ ($k=4)$, we can write the gamma matrices as
\begin{equation}
	\begin{split}
		\Gamma^\mu_{10} &=\Gamma^\mu_{4} \otimes \tilde\Gamma_{6}, \qquad \mu=0,\dots,3, \\
		\Gamma^4_{10} &=I_4 \otimes \tilde\Gamma^1_{6}, \\
		\Gamma^5_{10} &=I_4 \otimes \tilde\Gamma^0_{6}, \\
		\Gamma^p_{10} &=I_4 \otimes \tilde\Gamma^{p-4}_{6}, \qquad p=6,\dots,9.
	\end{split}
\end{equation}
We define $\tilde{\Gamma}^p$ with $p=1, \dots, 6$ as the reordered 6D Euclidean gamma matrices $\tilde\Gamma^1_{6}$, $\tilde\Gamma^0_{6}$, $\tilde\Gamma^2_{6}, \dots, \tilde\Gamma^5_{6}$, and we write $\gamma^{\mu} = \Gamma_4^{\mu}$ for the 4D matrices. We also write $\gamma_5=\Gamma_4$, $\tilde{\Gamma}= \tilde{\Gamma}_6$, and $\Gamma= \Gamma_{10}$ for the chirality matrices in 4D, 6D, and 10D, respectively. We can then write the gamma matrices and chirality matrix in 10D as
\begin{equation}
	\begin{split}
		\Gamma^\mu &=\gamma^\mu \otimes \tilde\Gamma, \qquad \mu=0,\dots,3, \\
		\Gamma^p &=I_4 \otimes \tilde\Gamma^{p}, \qquad p=1,\dots,6, \\
		\Gamma & =-\gamma_5 \otimes \tilde{\Gamma}.
	\end{split}
\end{equation}

\section{Example manifolds}\label{app:examples}

In this appendix, we show explicitly how the simplest sum rule is satisfied for specific manifolds. We consider flat tori and round spheres, for which the Dirac and scalar Laplace spectra are known analytically.

\subsection{Sum rule}

The sum rule we consider is the analogue of \eqref{eq:SRApsiApsi_2} for Dirac spinors. Let $\chi_i$ form an orthonormal basis of Dirac eigenspinors,\footnote{In terms of the definite chirality spinors $\chi_{\pm, i}$, we can form a basis of Dirac eigenspinors $\{ \chi_i^{ (\pm) } \}$, where $$  \chi_i^{(\pm)}  =\frac{1}{\sqrt{2}} \left( \chi_{+,i}\pm\chi_{-,i} \right).$$ }
\begin{equation}
	i\tilde\Gamma^p\nabla_p\chi_i =  \lambda_i\chi_i,
\end{equation}
and define the triple integrals $c'_{jic}$ by
\begin{equation}
	c'_{jic}\coloneq\int_\mathcal{M}\zeta_c\,\chi^\dagger_j\chi_i.
\end{equation}
We obtain a sum rule by evaluating  the quartic integral of $\zeta_a \zeta_b \chi^\dagger_j \chi_i$ in the following two ways:
\begin{equation}
	\int_\mathcal{M} \wick{
		\c1 \zeta_a \c1 \zeta_b \c2 \chi^\dagger_j \c2 \chi_i
	} =\int_\mathcal{M} \wick{
		\c1 \zeta_a \c2 \zeta_b \c1 \chi^\dagger_j \c2 \chi_i
	}.
\end{equation}
This gives the following sum rule, which holds for any dimension $n$ of the internal space:
\begin{equation}\label{eq:sumruleEquiv}
	0 =\frac{\delta_{ab} \delta_{ij}}{V} +\sum_c c_{abc} \, c'_{jic} -\sum_k c'_{jka} \, c'_{kib}.
\end{equation}
This has the same form as \eqref{eq:SRApsiApsi_2} and corresponds to the simplest unitarity sum rule from \cite{LlewellynSmith:1973yud, Gunion:1990kf, Ohl:2003dp, Schwinn:2004xa}. Taking identical external modes, i.e., $a=b$ and $i=j$, 
 the sum rule reduces to
\begin{equation} \label{eq:simple_SR_identical}
	0 =\frac1V +\sum_c c_{aac} \, c'_{iic} -\sum_k |c'_{ika}|^2.
\end{equation}

\subsection{Flat tori}\label{sec:exampletori}

Consider a flat torus $T^n\coloneq\mathbb{R}^n/\Lambda$, where $\Lambda$ is an $n$-dimensional full-rank Euclidean lattice with basis $\{\vec{v}_1,\dots,\vec{v}_n\}$ and a spin structure determined by an $n$-tuple $(\delta_1,\dots,\delta_n)$, with $\delta_i\in\{0,1\}$. The dual lattice of $\Lambda$ is denoted by $\Lambda^*$.

A real basis of eigenfunctions of the scalar Laplacian is given by
\begin{equation}
		\zeta_0=\frac1{\sqrt{V}}, \qquad \zeta^+_{\vec{k}} =\sqrt{\frac2{V}}\cos(2\pi \vec{k} \cdot \vec{y}), \qquad \zeta^-_{\vec{k}} =\sqrt{\frac2{V}}\sin(2\pi \vec{k} \cdot \vec{y}),
\end{equation}
where  $V$ is the volume of $T^n$ and $\vec{k}$ runs over the nonzero vectors in $\Lambda^*$ with $\vec{k} \sim - \vec{k}$. The constant zero mode is $\zeta_0$ and $\zeta^\pm_{\vec{k}}$ have eigenvalues $\lambda=4\pi^2|\vec{k}|^2$.

The spectrum of the Dirac operator on flat tori is given in \cite{Th1984} (see also \cite{diracSpect_book, baer2000}). To construct the eigenspinors on $T^n$, consider spinors of the form
\begin{equation}
	\chi_{\vec{k},\ell} \coloneq e^{2\pi i\, \vec{k}_\delta \cdot \vec{y}}\sigma_\ell,
\end{equation}
where $\sigma_\ell$ with $\ell=1,\dots,2^{\lfloor\frac{n}2\rfloor}$ is an orthonormal basis of spinors, $\sigma^{\dagger}_{\ell} \sigma_{\ell'} =\delta_{\ell\ell'}$, and we defined
\begin{equation}
	\vec{k}_\delta \coloneq \vec{k} +\frac12\sum_{i=1}^n \delta_i \pvec{v}_i^*, \qquad \vec{k}\in\Lambda^*,
\end{equation}
with $\{\pvec{v}^*_1,\dots,\pvec{v}^*_n\}$ the basis dual to $\{\vec{v}_1,\dots,\vec{v}_n\}$.
Acting with the Dirac operator
\begin{equation}
	\mathcal{D}^{(n)}\coloneq  i\tilde\Gamma^p \nabla_p,
\end{equation}
we get
\begin{equation} \label{eq:pre_Dirac_eigenspinor}
	\mathcal{D}^{(n)}\chi_{\vec{k},\ell} 
	=-2\pi  |\vec{k}_\delta| \, e^{2\pi i\, \vec{k}_\delta \cdot \vec{y}}({\tilde\Gamma}\cdot \hat{k}_\delta) \sigma_\ell,
\end{equation}
where $\hat{k}_\delta=\vec{k}_\delta/|\vec{k}_\delta|$ and ${\tilde\Gamma}\cdot \hat{k}_\delta = \tilde{\Gamma}^p (\hat{k}_\delta)^q \delta_{pq}$.

From \eqref{eq:pre_Dirac_eigenspinor}, we see that $\chi_{\vec{k},\ell} $ is a harmonic spinor if and only if $\vec{k}_\delta=0$. Such spinors exist only if the spin structure is the trivial one, $\delta_1=\ldots=\delta_n=0$, in which case there are $2^{\lfloor\frac{n}2\rfloor}$ of them. The $L^2$-normalized harmonic spinors are then given by
\begin{equation}
	\chi_{0,\ell} = \frac{1}{\sqrt{V}} \sigma_\ell.
\end{equation}
For non-harmonic eigenspinors, we need to diagonalize the matrix ${\tilde\Gamma}\cdot \hat{k}_\delta$. For any unit vector $\hat{n}$, the matrix $\tilde\Gamma\cdot \hat{n}$ has eigenvalues $\pm 1$, since $(\tilde\Gamma\cdot \hat{n})^2 =I_n$. For a given $\vec{k} \in \Lambda^*$, we can thus decompose the space of spinors as
\begin{equation}
	\mathrm{Ker}\left(\tilde\Gamma\cdot \hat{k}_\delta -I_n\right) \oplus \mathrm{Ker}\left(\tilde\Gamma\cdot \hat{k}_\delta +I_n\right).
\end{equation}
We thus replace the basis of spinors $(\sigma_1,\dots,\sigma_{2^{\lfloor\frac{n}2\rfloor}})$ with $(\sigma^+_1,\dots,\sigma^+_{2^{\lfloor\frac{n}2\rfloor-1}},\sigma^-_1,\dots,\sigma^-_{2^{\lfloor\frac{n}2\rfloor-1}})$, where $(\sigma^\varepsilon_1,\dots,\sigma^\varepsilon_{2^{\lfloor\frac{n}2\rfloor-1}})$ is an orthonormal basis of the space $\mathrm{Ker}\left( \tilde\Gamma\cdot \hat{k}_\delta -\varepsilon\,I_n\right)$ and $\varepsilon=\pm1$.
The basis satisfies
\begin{equation}\label{eq:toriDiracON}
	\begin{split}
		\tilde\Gamma\cdot \hat{k}_\delta\,\sigma^\varepsilon_\ell &=\varepsilon\,\sigma^\varepsilon_\ell, \\
		\sigma^{\varepsilon\dagger}_{\ell}\sigma^{\varepsilon'}_{\ell'} &=\delta_{\ell\ell'}\delta_{\varepsilon\varepsilon'}.
	\end{split}
\end{equation}
The orthonormal spinor eigenmodes of the Dirac operator with $\vec{k}_{\delta} \neq 0$ are thus given by
\begin{equation}\label{eq:toriDiracEigenspinors}
	\chi^\varepsilon_{\vec{k},\ell} \coloneq \frac{1}{\sqrt{V}} e^{2\pi i\, \vec{k}_\delta \cdot \vec{y}}\sigma^\varepsilon_\ell,
\end{equation}
which satisfy
\begin{equation}
	\mathcal{D}^{(n)}\chi^\varepsilon_{\vec{k},\ell} =-2\pi \varepsilon |\vec{k}_\delta| \chi^\varepsilon_{\vec{k},\ell}.
\end{equation}
The nonzero eigenvalues have multiplicities divisible by $2^{\lfloor\frac{n}2\rfloor-1}$.

We can now check \eqref{eq:simple_SR_identical}. We consider $T^6$ with trivial spin structure, i.e., $\delta_1=\ldots=\delta_6=0$. As external eigenmodes, we take $\zeta_a=\zeta^{\pm}_{\vec{k}_1}$ and $\chi_i=\chi^\varepsilon_{\vec{k}_2,\ell}$ with $\vec{k}_1, \vec{k}_2 \neq 0$.
By \eqref{eq:toriDiracON} and \eqref{eq:toriDiracEigenspinors}, we have
\begin{equation}
c'_{iic}= \int_{T^6} \zeta_c\chi^\dagger_i\chi_i =\frac1V\int_{T^6}\zeta_c =0,
\end{equation}
since $\zeta_c$ is non-constant. Writing $\chi_k=\chi_{\vec{k},\ell'}^{\varepsilon'}$, we also have
\begin{equation}
c_{ika}'=	\int_{T^6}\zeta_a \chi^\dagger_i\chi_k =\frac1{\sqrt{2V}}\frac1{i^{(1\mp1)/2}}\left(\delta_{\vec{k}_1 -\vec{k}_2+\vec{k} }\pm \delta_{\vec{k}_1 +\vec{k}_2-\vec{k} } \right)\sigma_{\ell}^{\varepsilon}(\hat{k}_2)^\dagger \sigma_{\ell'}^{\varepsilon'} (\hat{k}) .
\end{equation}
Using completeness in the form $\sum_{\varepsilon', \ell'} \sigma_{\ell'}^{\varepsilon'}  (\sigma_{\ell'}^{\varepsilon'})^{\dagger}=1$, this gives
\begin{align}
\sum_k |c'_{ika} |^2 & = \frac{1}{2V} \sum_{\vec{k} = \vec{k}_2 \pm \vec{k}_1} \sum_{\varepsilon', \ell'} \Big| (\sigma_{\ell}^{\varepsilon}(\hat{k}_2))^\dagger \sigma_{\ell'}^{\varepsilon'} (\hat{k}) \Big|^2  = \frac{1}{V}  (\sigma_{\ell}^{\varepsilon}(\hat{k}_2))^\dagger \sigma_{\ell}^{\varepsilon}(\hat{k}_2) = \frac{1}{V}.
\end{align}
We thus see explicitly that \eqref{eq:simple_SR_identical} is satisfied in this case. One can similarly show that the sum rule is satisfied for non-identical eigenmodes as well.

\subsection{Spheres}
We next consider eigenmodes on $n$-spheres, $\mathbb{S}^n$, which we always take to be equipped with the round metric and to have unit radius. We use the results of \cite{Higuchi:1986wu,Camporesi:1995fb, 199669} for the Laplace and Dirac eigenmodes on $\mathbb{S}^n$. To use the formulas from these references, we follow the conventions of \cite{Camporesi:1995fb} for the Euclidean gamma matrices in $n$ dimensions, which differ from the ones in Appendix \ref{app:clifford}.

In geodesic polar coordinates $(\theta_1 \coloneq\phi,\theta_2,\dots,\theta_n)$, the round metric is given by
\begin{equation}
	ds^2_n =d\theta^2_n +\sin^2\theta_n ds^2_{n-1},
\end{equation}
with $ds^2_1 =d\phi^2$, $0\leq\phi<2\pi$ and $0\leq\theta_n\leq\pi$. We therefore have
\begin{equation}
	\int_{\mathbb{S}^n} f(\phi,\theta_2,\dots,\theta_n) \coloneq \int_0^\pi d\theta_n\cdots \int_0^\pi d\theta_2 \int_0^{2\pi} d\phi\,(\sin\theta_n)^{n-1}(\sin\theta_{n-1})^{n-2}\cdots\sin\theta_2\,f(\phi,\theta_2,\dots,\theta_n).
\end{equation}
The volume of $\mathbb{S}^n$ is
\begin{equation}
	V_n=\frac{2\pi^{(n+1)/2}}{\Gamma\big((n+1)/2\big)}.
\end{equation}
For every $n\ge2$, the sphere $\mathbb{S}^n$ admits a unique spin structure.

\subsubsection{Laplace and Dirac eigenmodes}

Let us review the spectrum of the scalar Laplace operator  $\Delta^{(n)} \coloneq-\nabla^p\nabla_p$  on  $\mathbb{S}^n$ \cite{Berger1971}.
A basis of scalar eigenfunctions is given by the generalized spherical harmonics,
\begin{equation}
	\begin{split}
		&Y_{\ell_n,\dots,\ell_1}(\theta_n,\dots,\theta_2,\phi) \\ &\quad\quad\quad\quad =\frac{e^{i\ell_1\phi}}{\sqrt{2\pi}} \prod_{m=2}^n \sqrt{\frac{(2\ell_m+m-1)(\ell_{m-1}+\ell_m+m-2)!}{2(\ell_m-\ell_{m-1})!}}(\sin\theta_m)^{-\frac{(m-2)}2} P^{-\ell_{m-1}-\frac{(m-2)}2}_{\ell_m+\frac{(m-2)}2}(\cos\theta_m),
	\end{split}
\end{equation}
where $\ell_1,\dots,\ell_n$ are integers, $\ell_n\ge\ell_{n-1}\ge\cdots\ge\ell_2\ge|\ell_1|$ and $P^{a}_{b}(x)$ are associated Legendre functions of the first kind. These satisfy the eigenvalue equation
\begin{equation}
	\Delta^{(n)} Y_{\ell_n,\dots,\ell_1} =\ell_n(\ell_n+n-1)Y_{\ell_n,\dots,\ell_1}.
\end{equation}
The orthogonality condition reads
\begin{equation}
	\int_{\mathbb{S}^n} Y_{\ell'_n,\dots,\ell'_1}(\theta_n,\dots,\theta_2,\phi)^* Y_{\ell_n,\dots,\ell_1}(\theta_n,\dots,\theta_2,\phi) =\delta_{\ell_n\ell'_n}\cdots\delta_{\ell_1\ell'_1}.
\end{equation}
We can obtain a real orthonormal basis by taking
\begin{equation}
	\begin{split}
		\zeta_{\ell_n,\dots,\ell_1}(\theta_n,\dots,\theta_2,\phi) \coloneq\begin{cases}
			\frac1{\sqrt{2}}\big( Y_{\ell_n,\dots,-\ell_1} +(-1)^{\ell_1}Y_{\ell_n,\dots,\ell_1} \big) \qquad &\mathrm{if} \quad \ell_1>0, \\
			Y_{\ell_n,\dots,\ell_2,0} \qquad &\mathrm{if} \quad \ell_1=0, \\
			\frac{i}{\sqrt{2}}\big( Y_{\ell_n,\dots,\ell_1} -(-1)^{\ell_1}Y_{\ell_n,\dots,-\ell_1} \big) \qquad &\mathrm{if} \quad \ell_1<0.
		\end{cases}
	\end{split}
\end{equation}
To simplify calculations, we restrict to zonal spherical harmonics, with $\ell_{n-1}=\dots=\ell_1=0$:
\begin{equation}
	\zeta_{\ell_n}(\theta_n) \coloneq Y_{\ell_n,0,\dots,0}(\theta_n) =c_n(\ell_n)C_{\ell_n}^{\left(\frac{n-1}2\right)}(\cos\theta_n),
\end{equation}
where $C^{(a)}_b(x)$ are Gegenbauer polynomials and
\begin{equation}
	c_n(\ell_n) =\sqrt{\frac{(2\ell_n+n-1)\Gamma(n-1)\Gamma\big((n-1)/2\big)\Gamma(\ell_n+1)}{4\pi^{(n+1)/2}\Gamma(\ell_n+n-1)}}.
\end{equation}

The spectrum of the Dirac operator $ i\tilde\Gamma^p \nabla_p$ on $\mathbb{S}^n$ was computed in \cite{Camporesi:1995fb}. To find eigenspinors of the Dirac operator on $\mathbb{S}^n$, we make use of eigenspinors on $\mathbb{S}^{n-1}$.
For later purposes, we define the functions $\phi_{\ell\ell'}(\theta)$ and $\psi_{\ell\ell'}(\theta)$:
\begin{equation}
	\begin{split}
		\phi_{\ell\ell'}(\theta) &=\left(\cos \frac{\theta}2\right)^{\ell'+1}\left(\sin \frac{\theta}2\right)^{\ell'} P_{\ell-\ell'}^{\left(\frac{n}2+\ell'-1,\frac{n}2+\ell'\right)}(\cos\theta), \\
		\psi_{\ell\ell'}(\theta) &=\left(\cos \frac{\theta}2\right)^{\ell'}\left(\sin \frac{\theta}2\right)^{\ell'+1} P_{\ell-\ell'}^{\left(\frac{n}2+\ell',\frac{n}2+\ell'-1\right)}(\cos\theta),
	\end{split}
\end{equation}
where $P^{(a,b)}_\ell(x)$ are Jacobi polynomials. These are related by $\psi_{\ell\ell'}(\theta) =(-1)^{\ell-\ell'}\phi_{\ell\ell'}(\pi-\theta)$. We also define the normalization $c_n(\ell,\ell')$ such that
\begin{equation}
	|c_n(\ell,\ell')|^2 =\frac{(\ell-\ell')!(n+\ell+\ell'-1)!}{2^{n-2}\left|\Gamma(\frac{n}2+\ell)\right|^2}.
\end{equation}

\paragraph{Starting point: $n=1$.} Consider the circle $\mathbb{S}^1$. This has two spin structures; in the trivial case, the spectrum is given by the integers. In the non-trivial case, the Dirac eigenmodes satisfy
\begin{equation}
	i\frac{\partial}{\partial\phi}\psi_{\epsilon_1,\ell_1} =-\epsilon_1\left(\ell_1+\frac12\right)\psi_{\epsilon_1,\ell_1}
\end{equation}
and read
\begin{equation}
	\psi_{\epsilon_1,\ell_1}(\phi) =\frac1{\sqrt{2\pi}}e^{\epsilon_1 i(\ell_1+1/2)\phi},
\end{equation}
where $\epsilon_1=\pm$, $\ell_1=0,1,2, \dots$, and $0\le\phi<2\pi$.

\paragraph{$n$ even.}

In block diagonal form, we have
\begin{equation}
	\begin{split}
		&\psi^{(s_n,s_{n-2},s_{n-4},\dots, s_2)}_{\epsilon_n,\ell_n,\ell_{n-1},\dots,\ell_1}(\theta_n,\theta_{n-1},\dots,\theta_2,\phi) \\
		&\quad\quad\quad\quad\quad\quad\quad\quad =\frac{c_n(\ell_n,\ell_{n-1})}{\sqrt{2}}\begin{pmatrix}
			f^{(s_n)}_{\ell_n\ell_{n-1}}(\theta_n)\psi^{(s_{n-2},s_{n-4},\dots, s_2)}_{s_n,\ell_{n-1},\dots,\ell_1}(\theta_{n-1},\dots,\theta_2,\phi) \\
			\epsilon_n\,f^{(-s_n)}_{\ell_n\ell_{n-1}}(\theta_n)\psi^{(s_{n-2},s_{n-4},\dots, s_2)}_{s_n,\ell_{n-1},\dots,\ell_1}(\theta_{n-1},\dots,\theta_2,\phi)
		\end{pmatrix},
	\end{split}
\end{equation}
for $s_n=\pm$, $\epsilon_n=\pm$, $0\le\theta_n\le\pi$, $\ell_n=0,1,\dots$, and $\ell_n\ge \ell_{n-1}$. In this formula, $\psi^{(s_{n-2},s_{n-4},\dots, s_2)}_{s_n,\ell_{n-1},\dots,\ell_1}$ is an eigenmode in $n-1$ dimensions (when $n=2$ the superscript is absent) and we define
\begin{equation}
	\begin{split}
		f^{(-)}_{\ell_n\ell_{n-1}}(\theta_n)
		&=\phi_{\ell_n\ell_{n-1}}(\theta_n), \\
		f^{(+)}_{\ell_n\ell_{n-1}}(\theta_n)
		&=i\psi_{\ell_n\ell_{n-1}}(\theta_n).
	\end{split}
\end{equation}
These satisfy
\begin{equation}
	\mathcal{D}^{(n)}\psi^{(s_n,s_{n-2},\dots, s_2)}_{\epsilon_n,\ell_n,\ell_{n-1},\dots,\ell_1} =-\epsilon_n\left(\ell_n+\frac{n}2\right)\psi^{(s_n,s_{n-2},\dots,s_2)}_{\epsilon_n,\ell_n,\ell_{n-1},\dots,\ell_1}.
\end{equation}
From this, we see that there are no non-trivial harmonic spinors on an $n$-sphere.
The orthogonality condition reads
\begin{equation}
	\begin{split}
		&\int_{\mathbb{S}^n} \psi^{(s'_n,s'_{n-2},s'_{n-4},\dots, s'_2)}_{\epsilon'_n,\ell'_n,\ell'_{n-1},\dots,\ell'_1}(\theta_n,\theta_{n-1},\dots,\theta_2,\phi)^\dagger \psi^{(s_n,s_{n-2},s_{n-4},\dots, s_2)}_{\epsilon_n,\ell_n,\ell_{n-1},\dots,\ell_1}(\theta_n,\theta_{n-1},\dots,\theta_2,\phi) \\
		&\quad\quad\quad\quad\quad =\delta_{\ell_n\ell'_n}\cdots\delta_{\ell_1\ell'_1}\delta_{s_ns'_n}\cdots\delta_{s_2s'_2}\delta_{\epsilon_n\epsilon'_n}.
	\end{split}
\end{equation}

\paragraph{$n$ odd ($n\ge 3$).} In odd dimensions, we take the combination
\begin{equation}
	\begin{split}
		&\psi^{(s_{n-1},s_{n-3},\dots,s_2)}_{\epsilon_n,\ell_n,\ell_{n-1},\dots,\ell_1}(\theta_n,\theta_{n-1},\dots,\theta_2,\phi) \\
		&\quad\quad\quad\quad\quad =\frac{c_n(\ell_n,\ell_{n-1})}2\phi_{\ell_n\ell_{n-1}}(\theta_n)\Big(I+i\tilde\Gamma^n\Big)\psi^{(s_{n-1},s_{n-3},\dots,s_2)}_{-,\ell_{n-1},\ell_{n-2},\dots,\ell_1}(\theta_{n-1},\dots,\theta_2,\phi) \\
		&\quad\quad\quad\quad\quad\quad -\epsilon_n\frac{c_n(\ell_n,\ell_{n-1})}2\psi_{\ell_n\ell_{n-1}}(\theta_n)\Big(I-i\tilde\Gamma^n\Big)\psi^{(s_{n-1},s_{n-3},\dots,s_2)}_{-,\ell_{n-1},\ell_{n-2},\dots,\ell_1}(\theta_{n-1},\dots,\theta_2,\phi),
	\end{split}
\end{equation}
for $\epsilon_n=\pm$, $0\le\theta_n\le\pi$, $\ell_n=0,1,\dots$, and $\ell_n\ge \ell_{n-1}$. These satisfy
\begin{equation}
	\mathcal{D}^{(n)}\psi^{(s_{n-1},s_{n-3},\dots, s_2)}_{\epsilon_n,\ell_n,\ell_{n-1},\dots,\ell_1} =-\epsilon_n\left(\ell_n+\frac{n}2\right)\psi^{(s_{n-1},s_{n-3},\dots, s_2)}_{\epsilon_n,\ell_n,\ell_{n-1},\dots,\ell_1}.
\end{equation}
The orthogonality condition reads
\begin{equation}
	\begin{split}
		&\int_{\mathbb{S}^n} \psi^{(s'_{n-1},s'_{n-3},\dots,s'_2)}_{\epsilon'_n,\ell'_n,\ell'_{n-1},\dots,\ell'_1}(\theta_n,\theta_{n-1},\dots,\theta_2,\phi)^\dagger \psi^{(s_{n-1},s_{n-3},\dots,s_2)}_{\epsilon_n,\ell_n,\ell_{n-1},\dots,\ell_1}(\theta_n,\theta_{n-1},\dots,\theta_2,\phi) \\
		&\quad\quad\quad\quad\quad =\delta_{\ell_n\ell'_n}\cdots\delta_{\ell_1\ell'_1}\delta_{s_{n-1}s'_{n-1}}\cdots\delta_{s_2s'_2}\delta_{\epsilon_n\epsilon'_n}.
	\end{split}
\end{equation}

\subsubsection{Triple overlaps for zonal eigenfunctions}\label{sec:czonal}

The scalar triple product for zonal spherical harmonics,
\begin{equation}
	\int_{\mathbb{S}^n} \zeta_{\ell_n}(\theta_n)\zeta_{\ell'_n}(\theta_n)\zeta_{\ell''_n}(\theta_n),
\end{equation}
is non-zero only when $\ell_n+\ell'_n+\ell''_n$ is even and
\begin{equation}\label{eq:cabcTriangle}
	|\ell_n-\ell'_n|\le \ell''_n \le \ell_n+\ell'_n.
\end{equation}
When these conditions are satisfied, it reads
\begin{equation}
	\int_{\mathbb{S}^n} \zeta_{\ell_n}(\theta_n)\zeta_{\ell'_n}(\theta_n)\zeta_{\ell''_n}(\theta_n) =c_n(\ell_n)c_n(\ell'_n)c_n(\ell''_n) \frac{2\pi^{n/2}}{\Gamma(n/2)} B^{(n-1)/2}_{\ell_n,\ell'_n,\ell''_n}\,h^{(n-1)/2}_{\ell''_n},
\end{equation}
where
\begin{equation}
	\begin{split}
		B^{\lambda}_{\ell_1,\ell_2,\ell_3} &=\frac{2(\ell_3+\lambda)\,\ell_3!(\lambda)_{\frac{\ell_1+\ell_2-\ell_3}{2}}(\lambda)_{\frac{\ell_1-\ell_2+\ell_3}{2}}(\lambda)_{\frac{-\ell_1+\ell_2+\ell_3}{2}}(2\lambda)_{\frac{\ell_1+\ell_2+\ell_3}{2}}}{(\ell_1+\ell_2+\ell_3+2\lambda)\big(\frac{\ell_1+\ell_2-\ell_3}{2}\big)!\big(\frac{\ell_1-\ell_2+\ell_3}{2}\big)!\big(\frac{-\ell_1+\ell_2+\ell_3}{2}\big)!(\lambda)_{\frac{\ell_1+\ell_2+\ell_3}{2}}(2\lambda)_{\ell_3}
		}, \\
		h^{\lambda}_\ell &=\frac{\pi  2^{1-2 \lambda } \Gamma (\ell+2 \lambda )}{\ell! (\ell+\lambda ) \Gamma (\lambda )^2}.
	\end{split}
\end{equation}

To study the fermion scalar overlaps $c'_{jic}$, we take $n$ to be even. The triple overlap with a non-constant zonal eigenfunction is
\begin{align}
		& \int_{\mathbb{S}^n} \zeta_{\ell''_n}(\theta_n) \psi^{(s'_n,s'_{n-2},s'_{n-4},\dots, s'_2)}_{\epsilon'_n,\ell'_n,\ell'_{n-1},\dots,\ell'_1}(\theta_n,\dots,\theta_2,\phi)^\dagger\psi^{(s_n,s_{n-2},s_{n-4},\dots, s_2)}_{\epsilon_n,\ell_n,\ell_{n-1},\dots,\ell_1}(\theta_n,\dots,\theta_2,\phi) \nonumber \\
		&\quad =\frac{c_n(\ell_n,\ell_{n-1})c_n(\ell'_n,\ell_{n-1})}2 c_n(\ell''_n)\, \delta_{\ell_{n-1}\ell'_{n-1}}\cdots\delta_{\ell_1\ell'_1}\delta_{s_ns'_n}\delta_{s_{n-2}s'_{n-2}}\cdots\delta_{s_2s'_2}\,I^{(s_n)}_{\epsilon_n\epsilon'_n}(\ell_n,\ell'_n,\ell''_n), \label{eq:sphere_spinor_overlap}
		\end{align}
where we have defined
\begin{equation}
	I^{(s_n)}_{\epsilon_n\epsilon'_n}(\ell_n,\ell'_n,\ell''_n) =\int_0^\pi d\theta_n (\sin\theta_n)^{n-1} C_{\ell''_n}^{\left(\frac{n-1}2\right)}(\cos\theta_n) \Big[ f^{(s_n)*}_{\ell'_n\ell_{n-1}}(\theta_n)f^{(s_n)}_{\ell_n\ell_{n-1}}(\theta_n) +\epsilon'_n\epsilon_n\,f^{(-s_n)*}_{\ell'_n\ell_{n-1}}(\theta_n)f^{(-s_n)}_{\ell_n\ell_{n-1}}(\theta_n) \Big].
\end{equation}
These integrals evaluate to
\footnotesize
\begin{align}
		&I^{(-)}_{\epsilon_n\epsilon'_n}(\ell_n,\ell'_n,\ell''_n)=\frac{2^{n-1} B(Z,Z+1) (Z+1)_\alpha (Z+1)_\beta (n-1)_\gamma}{\alpha! \beta! \gamma!}\sum_{r=0}^{\alpha}\sum_{s=0}^{\beta}\sum_{k=0}^{\gamma} \frac{(-\alpha)_r (-\beta)_s (-\gamma)_k (\alpha+2 Z)_r (2 Z+\beta)_s (\gamma+n-1)_k}{k! r! s! \left(\frac{n}{2}\right)_k (2Z+1)_{k+r+s}} \nonumber \\
		&\,\,\,\,\times\frac{Z\, \Gamma (Z) \Gamma (Z+k+r+s)}{\Gamma (Z+r) \Gamma (Z+s)}\bigg[\frac{Z}{(\alpha+Z) (Z+\beta)}+\epsilon_n\epsilon'_n\frac{(Z+k+r+s)}{(Z+r) (Z+s)}\bigg] \label{eq:Iminus}
\end{align}
\normalsize
and
\begin{equation}
	I^{(+)}_{\epsilon_n\epsilon'_n}(\ell_n,\ell'_n,\ell''_n) =\epsilon_n\epsilon'_n I^{(-)}_{\epsilon_n\epsilon'_n}(\ell_n,\ell'_n,\ell''_n),
\end{equation}
where $B(z_1,z_2)$ is the Euler beta function, $\alpha=\ell_n-\ell_{n-1}$, $\beta=\ell'_n-\ell_{n-1}$, $\gamma=\ell''_n$, and $Z=\ell_{n-1}+\frac{n}2$.
In both cases, we have
\begin{equation}\label{eq:Izero}
	I^{(s_n)}_{\epsilon_n\epsilon'_n}(\ell_n,\ell'_n,\ell''_n) =0 \quad \mathrm{when} \quad |\ell_n-\ell'_n|>\ell''_n \quad \mathrm{or} \quad \ell''_n>\ell_n+\ell'_n+1.
\end{equation}
When $\epsilon_n=\epsilon'_n$, the dependence on $s_n$ drops out and
\begin{equation}
	I^{(+)}_{\epsilon_n\epsilon_n}(\ell_n,\ell'_n,\ell''_n) =I^{(-)}_{\epsilon_n\epsilon_n}(\ell_n,\ell'_n,\ell''_n) \eqcolon I(\ell_n,\ell'_n,\ell''_n).
\end{equation}

\subsubsection{Sum rule check}

Now we check the sum rule \eqref{eq:simple_SR_identical}. We take $\zeta_a= \zeta_{\ell''_n}(\theta_n)$ and $$\chi_i =\psi^{(s_n,s_{n-2},s_{n-4},\dots, s_2)}_{\epsilon_n,\ell_n,\ell_{n-1},\dots,\ell_1}(\theta_n,\dots,\theta_2,\phi). $$
We sum over exchanged zonal scalar eigenmodes $\zeta_c= \zeta_{\ell''_{\rm exc}}(\theta_n)$ with $\ell''_{\rm exc}\in\{2,4,\dots,2\ell''_n\}$, due to \eqref{eq:cabcTriangle}. Non-zonal scalar eigenmodes do not contribute to the sum. By \eqref{eq:sphere_spinor_overlap}, we sum over exchanged spinor eigenmodes $$\chi_k =\psi^{(s_n,s_{n-2},s_{n-4},\dots, s_2)}_{\epsilon'_{\rm exc} ,\ell'_{\rm exc},\ell_{n-1},\dots,\ell_1}(\theta_n,\dots,\theta_2,\phi),$$ with $\ell'_{\rm exc}\ge\ell_{n-1}$ and $\epsilon'_{\rm exc}=\pm$. By \eqref{eq:Izero}, we only need to sum over $\ell'_{\rm exc}$ such that $|\ell_n-\ell'_{\rm exc}|\le\ell''_n$. 

Putting everything together, the sum rule for identical externals on $\mathbb{S}^n$ with $n$ even reads
\begin{equation}\label{eq:sumruleSphere}
	\begin{split}
		0 &=\frac{\Gamma\big((n+1)/2\big)}{2\pi^{(n+1)/2}} +\frac{\big(c_n(\ell_n,\ell_{n-1})c_n(\ell''_n)\big)^2}2 \sum_{\substack{\ell''_{\rm exc}=2 \\ \ell''_{\rm exc} \text{ even }}}^{2\ell''_n} B^{(n-1)/2}_{\ell''_n,\ell''_n,\ell''_{\rm exc}}\, I(\ell_n,\ell_n,\ell''_{\rm exc}) \\
		&\quad\quad 	-\frac{\big(c_n(\ell_n,\ell_{n-1})c_n(\ell''_n)\big)^2}4 \sum_{\epsilon'_{\rm exc}=\pm}\,\sum_{\ell'_{\rm exc}=\ell_{n-1}}^{\ell_n+\ell''_n}\Big| c_n(\ell'_{\rm exc},\ell_{n-1}) \,I^{(s_n)}_{\epsilon_n\epsilon'_{\rm exc}}(\ell_n,\ell'_{\rm exc},\ell''_n) \Big|^2.
	\end{split}
\end{equation}
This must be satisfied for any choice of positive integers $\ell''_n$, $s_n=\pm$, $\epsilon_n=\pm$, and non-negative integers $\ell_n$ and $\ell_{n-1}$, with $\ell_n\ge\ell_{n-1}$. We have verified in \texttt{Mathematica} that \eqref{eq:sumruleSphere} is satisfied for $n=2,4$, and $6$ with many different choices of the external parameters. 
We have similarly checked the sum rule \eqref{eq:sumruleEquiv} for various choices of non-identical fixed eigenmodes in $n=2,4$, and $6$.

\renewcommand{\em}{}
\bibliographystyle{utphys}
\addcontentsline{toc}{section}{References}
\bibliography{SYM-refs}

\providecommand{\href}[2]{#2}\begingroup\raggedright\begin{thebibliography}{10}

\bibitem{Kaluza:1921tu}
T.~Kaluza, ``{Zum Unit{\"a}tsproblem der Physik},''
  \href{http://dx.doi.org/10.1142/S0218271818700017}{{\em Sitzungsber. Preuss.
  Akad. Wiss. Berlin (Math. Phys. )} {\bf 1921} (1921)  966--972},
  \href{http://arxiv.org/abs/1803.08616}{{\tt arXiv:1803.08616
  [physics.hist-ph]}}.

\bibitem{Klein:1926tv}
O.~Klein, ``{Quantum Theory and Five-Dimensional Theory of Relativity. (In
  German and English)},'' \href{http://dx.doi.org/10.1007/BF01397481}{{\em Z.
  Phys.} {\bf 37} (1926)  895--906}.

\bibitem{Bailin:1987jd}
D.~Bailin and A.~Love, ``{Kaluza-Klein Theories},''
  \href{http://dx.doi.org/10.1088/0034-4885/50/9/001}{{\em Rept. Prog. Phys.}
  {\bf 50} (1987)  1087--1170}.

\bibitem{Duff:1986hr}
M.~J. Duff, B.~E.~W. Nilsson, and C.~N. Pope, ``{Kaluza-Klein Supergravity},''
  \href{http://dx.doi.org/10.1016/0370-1573(86)90163-8}{{\em Phys. Rept.} {\bf
  130} (1986)  1--142}.

\bibitem{Freund:1980xh}
P.~G.~O. Freund and M.~A. Rubin, ``{Dynamics of Dimensional Reduction},''
  \href{http://dx.doi.org/10.1016/0370-2693(80)90590-0}{{\em Phys. Lett. B}
  {\bf 97} (1980)  233--235}.

\bibitem{Witten:1981me}
E.~Witten, ``{Search for a Realistic Kaluza-Klein Theory},''
  \href{http://dx.doi.org/10.1016/0550-3213(81)90021-3}{{\em Nucl. Phys. B}
  {\bf 186} (1981)  412}.

\bibitem{Salam:1981xd}
A.~Salam and J.~A. Strathdee, ``{On Kaluza-Klein Theory},''
  \href{http://dx.doi.org/10.1016/0003-4916(82)90291-3}{{\em Annals Phys.} {\bf
  141} (1982)  316--352}.

\bibitem{Candelas:1985en}
P.~Candelas, G.~T. Horowitz, A.~Strominger, and E.~Witten, ``{Vacuum
  configurations for superstrings},''
  \href{http://dx.doi.org/10.1016/0550-3213(85)90602-9}{{\em Nucl. Phys. B}
  {\bf 258} (1985)  46--74}.

\bibitem{Chivukula:2001esy}
R.~S. Chivukula, D.~A. Dicus, and H.-J. He, ``{Unitarity of compactified
  five-dimensional Yang-Mills theory},''
  \href{http://dx.doi.org/10.1016/S0370-2693(01)01435-6}{{\em Phys. Lett. B}
  {\bf 525} (2002)  175--182}, \href{http://arxiv.org/abs/hep-ph/0111016}{{\tt
  arXiv:hep-ph/0111016}}.

\bibitem{Csaki:2003dt}
C.~Csaki, C.~Grojean, H.~Murayama, L.~Pilo, and J.~Terning, ``{Gauge theories
  on an interval: Unitarity without a Higgs},''
  \href{http://dx.doi.org/10.1103/PhysRevD.69.055006}{{\em Phys. Rev. D} {\bf
  69} (2004)  055006}, \href{http://arxiv.org/abs/hep-ph/0305237}{{\tt
  arXiv:hep-ph/0305237}}.

\bibitem{Ohl:2003dp}
T.~Ohl and C.~Schwinn, ``{Unitarity, BRST symmetry and Ward identities in
  orbifold gauge theories},''
  \href{http://dx.doi.org/10.1103/PhysRevD.70.045019}{{\em Phys. Rev. D} {\bf
  70} (2004)  045019}, \href{http://arxiv.org/abs/hep-ph/0312263}{{\tt
  arXiv:hep-ph/0312263}}.

\bibitem{Schwinn:2004xa}
C.~Schwinn, ``{Higgsless fermion masses and unitarity},''
  \href{http://dx.doi.org/10.1103/PhysRevD.69.116005}{{\em Phys. Rev. D} {\bf
  69} (2004)  116005}, \href{http://arxiv.org/abs/hep-ph/0402118}{{\tt
  arXiv:hep-ph/0402118}}.

\bibitem{Schwinn:2005qa}
C.~Schwinn, ``{Unitarity constraints on top quark signatures of Higgsless
  models},'' \href{http://dx.doi.org/10.1103/PhysRevD.71.113005}{{\em Phys.
  Rev. D} {\bf 71} (2005)  113005},
  \href{http://arxiv.org/abs/hep-ph/0504240}{{\tt arXiv:hep-ph/0504240}}.

\bibitem{Bonifacio:2019ioc}
J.~Bonifacio and K.~Hinterbichler, ``{Unitarization from Geometry},''
  \href{http://dx.doi.org/10.1007/JHEP12(2019)165}{{\em JHEP} {\bf 12} (2019)
  165}, \href{http://arxiv.org/abs/1910.04767}{{\tt arXiv:1910.04767
  [hep-th]}}.

\bibitem{LlewellynSmith:1973yud}
C.~H. Llewellyn~Smith, ``{High-Energy Behavior and Gauge Symmetry},''
  \href{http://dx.doi.org/10.1016/0370-2693(73)90692-8}{{\em Phys. Lett. B}
  {\bf 46} (1973)  233--236}.

\bibitem{Cornwall:1973tb}
J.~M. Cornwall, D.~N. Levin, and G.~Tiktopoulos, ``{Uniqueness of spontaneously
  broken gauge theories},''
  \href{http://dx.doi.org/10.1103/PhysRevLett.30.1268}{{\em Phys. Rev. Lett.}
  {\bf 30} (1973)  1268--1270}. [Erratum: Phys.Rev.Lett. 31, 572 (1973)].

\bibitem{Gunion:1990kf}
J.~F. Gunion, H.~E. Haber, and J.~Wudka, ``{Sum rules for Higgs bosons},''
  \href{http://dx.doi.org/10.1103/PhysRevD.43.904}{{\em Phys. Rev. D} {\bf 43}
  (1991)  904--912}.

\bibitem{Hinterbichler:2013kwa}
K.~Hinterbichler, J.~Levin, and C.~Zukowski, ``{Kaluza-Klein Towers on General
  Manifolds},'' \href{http://dx.doi.org/10.1103/PhysRevD.89.086007}{{\em Phys.
  Rev. D} {\bf 89} (2014) no.~8, 086007},
  \href{http://arxiv.org/abs/1310.6353}{{\tt arXiv:1310.6353 [hep-th]}}.

\bibitem{SekharChivukula:2019yul}
R.~Sekhar~Chivukula, D.~Foren, K.~A. Mohan, D.~Sengupta, and E.~H. Simmons,
  ``{Scattering amplitudes of massive spin-2 Kaluza-Klein states grow only as
  ${\cal O}(s)$},'' \href{http://dx.doi.org/10.1103/PhysRevD.101.055013}{{\em
  Phys. Rev. D} {\bf 101} (2020) no.~5, 055013},
  \href{http://arxiv.org/abs/1906.11098}{{\tt arXiv:1906.11098 [hep-ph]}}.

\bibitem{Chivukula:2020hvi}
R.~S. Chivukula, D.~Foren, K.~A. Mohan, D.~Sengupta, and E.~H. Simmons,
  ``{Massive Spin-2 Scattering Amplitudes in Extra-Dimensional Theories},''
  \href{http://dx.doi.org/10.1103/PhysRevD.101.075013}{{\em Phys. Rev. D} {\bf
  101} (2020) no.~7, 075013}, \href{http://arxiv.org/abs/2002.12458}{{\tt
  arXiv:2002.12458 [hep-ph]}}.

\bibitem{Hang:2021fmp}
Y.-F. Hang and H.-J. He, ``{Structure of Kaluza-Klein graviton scattering
  amplitudes from the gravitational equivalence theorem and double copy},''
  \href{http://dx.doi.org/10.1103/PhysRevD.105.084005}{{\em Phys. Rev. D} {\bf
  105} (2022) no.~8, 084005}, \href{http://arxiv.org/abs/2106.04568}{{\tt
  arXiv:2106.04568 [hep-th]}}.

\bibitem{Arkani-Hamed:2002bjr}
N.~Arkani-Hamed, H.~Georgi, and M.~D. Schwartz, ``{Effective field theory for
  massive gravitons and gravity in theory space},''
  \href{http://dx.doi.org/10.1016/S0003-4916(03)00068-X}{{\em Annals Phys.}
  {\bf 305} (2003)  96--118}, \href{http://arxiv.org/abs/hep-th/0210184}{{\tt
  arXiv:hep-th/0210184}}.

\bibitem{Schwartz:2003vj}
M.~D. Schwartz, ``{Constructing gravitational dimensions},''
  \href{http://dx.doi.org/10.1103/PhysRevD.68.024029}{{\em Phys. Rev. D} {\bf
  68} (2003)  024029}, \href{http://arxiv.org/abs/hep-th/0303114}{{\tt
  arXiv:hep-th/0303114}}.

\bibitem{deRham:2010ik}
C.~de~Rham and G.~Gabadadze, ``{Generalization of the Fierz-Pauli Action},''
  \href{http://dx.doi.org/10.1103/PhysRevD.82.044020}{{\em Phys. Rev. D} {\bf
  82} (2010)  044020}, \href{http://arxiv.org/abs/1007.0443}{{\tt
  arXiv:1007.0443 [hep-th]}}.

\bibitem{Bonifacio:2019mgk}
J.~Bonifacio, K.~Hinterbichler, and R.~A. Rosen, ``{Constraints on a
  gravitational Higgs mechanism},''
  \href{http://dx.doi.org/10.1103/PhysRevD.100.084017}{{\em Phys. Rev. D} {\bf
  100} (2019) no.~8, 084017}, \href{http://arxiv.org/abs/1903.09643}{{\tt
  arXiv:1903.09643 [hep-th]}}.

\bibitem{SekharChivukula:2019qih}
R.~Sekhar~Chivukula, D.~Foren, K.~A. Mohan, D.~Sengupta, and E.~H. Simmons,
  ``{Sum Rules for Massive Spin-2 Kaluza-Klein Elastic Scattering
  Amplitudes},'' \href{http://dx.doi.org/10.1103/PhysRevD.100.115033}{{\em
  Phys. Rev. D} {\bf 100} (2019) no.~11, 115033},
  \href{http://arxiv.org/abs/1910.06159}{{\tt arXiv:1910.06159 [hep-ph]}}.

\bibitem{Chivukula:2022kju}
R.~S. Chivukula, E.~H. Simmons, and X.~Wang, ``{Supersymmetry and sum rules in
  the Goldberger-Wise model},''
  \href{http://dx.doi.org/10.1103/PhysRevD.106.035026}{{\em Phys. Rev. D} {\bf
  106} (2022) no.~3, 035026}, \href{http://arxiv.org/abs/2207.02887}{{\tt
  arXiv:2207.02887 [hep-ph]}}.

\bibitem{deGiorgi:2023mdy}
A.~de~Giorgi and S.~Vogl, ``{Gravity-matter sum rules in models with a single
  extra-dimension},'' \href{http://dx.doi.org/10.1007/JHEP05(2024)315}{{\em
  JHEP} {\bf 05} (2024)  315}, \href{http://arxiv.org/abs/2311.01507}{{\tt
  arXiv:2311.01507 [hep-ph]}}.

\bibitem{Belavin:1984vu}
A.~A. Belavin, A.~M. Polyakov, and A.~B. Zamolodchikov, ``{Infinite Conformal
  Symmetry in Two-Dimensional Quantum Field Theory},''
  \href{http://dx.doi.org/10.1016/0550-3213(84)90052-X}{{\em Nucl. Phys. B}
  {\bf 241} (1984)  333--380}.

\bibitem{Rattazzi:2008pe}
R.~Rattazzi, V.~S. Rychkov, E.~Tonni, and A.~Vichi, ``{Bounding scalar operator
  dimensions in 4D CFT},''
  \href{http://dx.doi.org/10.1088/1126-6708/2008/12/031}{{\em JHEP} {\bf 12}
  (2008)  031},
\href{http://arxiv.org/abs/0807.0004}{{\tt arXiv:0807.0004 [hep-th]}}.

\bibitem{Poland:2018epd}
D.~Poland, S.~Rychkov, and A.~Vichi, ``{The Conformal Bootstrap: Theory,
  Numerical Techniques, and Applications},''
  \href{http://dx.doi.org/10.1103/RevModPhys.91.015002}{{\em Rev. Mod. Phys.}
  {\bf 91} (2019)  015002}, \href{http://arxiv.org/abs/1805.04405}{{\tt
  arXiv:1805.04405 [hep-th]}}.

\bibitem{Bonifacio:2020xoc}
J.~Bonifacio and K.~Hinterbichler, ``{Bootstrap Bounds on Closed Einstein
  Manifolds},'' \href{http://dx.doi.org/10.1007/JHEP10(2020)069}{{\em JHEP}
  {\bf 10} (2020)  069}, \href{http://arxiv.org/abs/2007.10337}{{\tt
  arXiv:2007.10337 [hep-th]}}.

\bibitem{Bonifacio:2021msa}
J.~Bonifacio, ``{Bootstrap bounds on closed hyperbolic manifolds},''
  \href{http://dx.doi.org/10.1007/JHEP02(2022)025}{{\em JHEP} {\bf 02} (2022)
  025}, \href{http://arxiv.org/abs/2107.09674}{{\tt arXiv:2107.09674
  [hep-th]}}.

\bibitem{Bonifacio:2021aqf}
J.~Bonifacio, ``{Bootstrapping closed hyperbolic surfaces},''
  \href{http://dx.doi.org/10.1007/JHEP03(2022)093}{{\em JHEP} {\bf 03} (2022)
  093}, \href{http://arxiv.org/abs/2111.13215}{{\tt arXiv:2111.13215
  [hep-th]}}.

\bibitem{Kravchuk:2021akc}
P.~Kravchuk, D.~Mazac, and S.~Pal, ``{Automorphic spectra and the conformal
  bootstrap},'' \href{http://dx.doi.org/10.1090/cams/26}{{\em Commun. Am. Math.
  Soc.} {\bf 4} (2024) no.~1, 1--63},
  \href{http://arxiv.org/abs/2111.12716}{{\tt arXiv:2111.12716 [hep-th]}}.

\bibitem{Radcliffe:2024jcg}
A.~Radcliffe, ``{Non-saturation of bootstrap bounds by hyperbolic orbifolds},''
  \href{http://dx.doi.org/10.1007/JHEP12(2025)115}{{\em JHEP} {\bf 12} (2025)
  115}, \href{http://arxiv.org/abs/2404.14479}{{\tt arXiv:2404.14479
  [hep-th]}}.

\bibitem{Bonifacio:2023ban}
J.~Bonifacio, D.~Mazac, and S.~Pal, ``{Spectral Bounds on Hyperbolic
  3-Manifolds: Associativity and the Trace Formula},''
  \href{http://dx.doi.org/10.1007/s00220-024-05222-0}{{\em Commun. Math. Phys.}
  {\bf 406} (2025)  51}, \href{http://arxiv.org/abs/2308.11174}{{\tt
  arXiv:2308.11174 [math.SP]}}.

\bibitem{Gesteau:2023brw}
E.~Gesteau, S.~Pal, D.~Simmons-Duffin, and Y.~Xu, ``{Bounds on spectral gaps of
  Hyperbolic spin surfaces},''
  \href{http://dx.doi.org/10.56994/JAMR.003.001.003}{{\em J. Assoc. Math. Res.}
  {\bf 3} (2025) no.~1, 72--139}, \href{http://arxiv.org/abs/2311.13330}{{\tt
  arXiv:2311.13330 [math.SP]}}.

\bibitem{Adve:2025rvf}
A.~Adve, J.~Bonifacio, P.~Kravchuk, D.~Mazac, S.~Pal, A.~Radcliffe, and
  G.~Rogelberg, ``{Weyl bound for trilinear periods via conformal bootstrap},''
  \href{http://arxiv.org/abs/2508.20576}{{\tt arXiv:2508.20576 [math.NT]}}.

\bibitem{Bertucci:2026fjh}
F.~Bertucci and J.~Bonifacio, ``{Bootstrapping Euclidean Lattices},''
  \href{http://arxiv.org/abs/2606.13771}{{\tt arXiv:2606.13771 [hep-th]}}.

\bibitem{Gliozzi:1976qd}
F.~Gliozzi, J.~Scherk, and D.~I. Olive, ``{Supersymmetry, Supergravity Theories
  and the Dual Spinor Model},''
  \href{http://dx.doi.org/10.1016/0550-3213(77)90206-1}{{\em Nucl. Phys. B}
  {\bf 122} (1977)  253--290}.

\bibitem{Polchinski_1998}
J.~Polchinski, {\em String Theory}.
\newblock Cambridge Monographs on Mathematical Physics. Cambridge University
  Press, 1998.

\bibitem{Lichnerowicz:1963}
A.~Lichnerowicz, ``{Spineurs harmoniques},'' {\em C. R. Acad. Sci. Paris} {\bf
  257} (1963)  7--9.

\bibitem{baer2000}
C.~B\"ar, ``{Dependence on the spin structure of the Dirac spectrum},''
  \href{http://arxiv.org/abs/math/0007131}{{\tt arXiv:math/0007131 [math.DG]}}.

\bibitem{Atiyah:1963zz}
M.~F. Atiyah and I.~M. Singer, ``{The index of elliptic operators on compact
  manifolds},'' \href{http://dx.doi.org/10.1090/S0002-9904-1963-10957-X}{{\em
  Bull. Am. Math. Soc.} {\bf 69} (1963)  422--433}.

\bibitem{Eguchi:1980jx}
T.~Eguchi, P.~B. Gilkey, and A.~J. Hanson, ``{Gravitation, Gauge Theories and
  Differential Geometry},''
  \href{http://dx.doi.org/10.1016/0370-1573(80)90130-1}{{\em Phys. Rept.} {\bf
  66} (1980)  213}.

\bibitem{Green_Schwarz_Witten_2012}
M.~B. Green, J.~H. Schwarz, and E.~Witten, {\em Superstring Theory: 25th
  Anniversary Edition}.
\newblock Cambridge Monographs on Mathematical Physics. Cambridge University
  Press, 2012.

\bibitem{Witten:1983ux}
E.~Witten, ``{Fermion Quantum Numbers in Kaluza-Klein Theory},'' {\em Conf.
  Proc. C} {\bf 8306011} (1983)  227.

\bibitem{Brink:1976bc}
L.~Brink, J.~H. Schwarz, and J.~Scherk, ``{Supersymmetric Yang-Mills
  Theories},'' \href{http://dx.doi.org/10.1016/0550-3213(77)90328-5}{{\em Nucl.
  Phys. B} {\bf 121} (1977)  77--92}.

\bibitem{Blau:2000xg}
M.~Blau, ``{Killing spinors and SYM on curved spaces},''
  \href{http://dx.doi.org/10.1088/1126-6708/2000/11/023}{{\em JHEP} {\bf 11}
  (2000)  023}, \href{http://arxiv.org/abs/hep-th/0005098}{{\tt
  arXiv:hep-th/0005098}}.

\bibitem{Festuccia:2011ws}
G.~Festuccia and N.~Seiberg, ``{Rigid Supersymmetric Theories in Curved
  Superspace},'' \href{http://dx.doi.org/10.1007/JHEP06(2011)114}{{\em JHEP}
  {\bf 06} (2011)  114}, \href{http://arxiv.org/abs/1105.0689}{{\tt
  arXiv:1105.0689 [hep-th]}}.

\bibitem{Duff:1982ev}
M.~J. Duff, ``{Supergravity, the Seven Sphere, and Spontaneous Symmetry
  Breaking},'' \href{http://dx.doi.org/10.1016/0550-3213(83)90649-1}{{\em Nucl.
  Phys. B} {\bf 219} (1983)  389}.

\bibitem{Awada:1982pk}
M.~A. Awada, M.~J. Duff, and C.~N. Pope, ``{N = 8 Supergravity Breaks Down to N
  = 1},'' \href{http://dx.doi.org/10.1103/PhysRevLett.50.294}{{\em Phys. Rev.
  Lett.} {\bf 50} (1983)  294}.

\bibitem{Duff:1983vj}
M.~J. Duff, B.~E.~W. Nilsson, and C.~N. Pope, ``{Compactification of d = 11
  Supergravity On K3 {\texttimes} T$^3$},''
  \href{http://dx.doi.org/10.1016/0370-2693(83)90724-4}{{\em Phys. Lett. B}
  {\bf 129} (1983)  39}.

\bibitem{Duff:1983gq}
M.~J. Duff and C.~N. Pope, ``{Kaluza-Klein supergravity and the
  seven-sphere},'' in {\em {September School on Supergravity and
  Supersymmetry}}.
\newblock 1983.

\bibitem{Duff:2025tot}
M.~J. Duff, B.~E.~W. Nilsson, and C.~N. Pope, ``{Kaluza-Klein Supergravity
  2025},'' \href{http://arxiv.org/abs/2502.07710}{{\tt arXiv:2502.07710
  [hep-th]}}.

\bibitem{HITCHIN19741}
N.~Hitchin, ``{Harmonic Spinors},''
  \href{http://dx.doi.org/https://doi.org/10.1016/0001-8708(74)90021-8}{{\em
  Advances in Mathematics} {\bf 14} (1974) no.~1, 1--55}.

\bibitem{Gubser:2002mz}
S.~S. Gubser, ``{TASI lectures: Special holonomy in string theory and M
  theory},'' \href{http://arxiv.org/abs/hep-th/0201114}{{\tt
  arXiv:hep-th/0201114}}.

\bibitem{10.1063/1.532293}
B.~McInnes, ``{Existence of parallel spinors on nonsimply connected Riemannian
  manifolds},'' \href{http://dx.doi.org/10.1063/1.532293}{{\em Journal of
  Mathematical Physics} {\bf 39} (1998) no.~4, 2362--2366}.

\bibitem{Wang1989}
M.~Y. Wang, ``{Parallel spinors and parallel forms},''
  \href{http://dx.doi.org/10.1007/BF00137402}{{\em Annals of Global Analysis
  and Geometry} {\bf 7} (1989)  59--68}.

\bibitem{Wang1995}
M.~Y. Wang, ``{On non-simply connected manifolds with non-trivial parallel
  spinors},'' \href{http://dx.doi.org/10.1007/BF00774565}{{\em Annals of Global
  Analysis and Geometry} {\bf 13} (1995)  31--42}.

\bibitem{Moroianu_2000}
A.~Moroianu and U.~Semmelmann, ``{Parallel spinors and holonomy groups},''
  \href{http://dx.doi.org/10.1063/1.533247}{{\em Journal of Mathematical
  Physics} {\bf 41} (2000) no.~4, 2395–2402},
  \href{http://arxiv.org/abs/math/9903062}{{\tt arXiv:math/9903062 [math.DG]}}.

\bibitem{Srednicki:2007qs}
M.~Srednicki, \href{http://dx.doi.org/10.1017/CBO9780511813917}{{\em {Quantum
  field theory}}}.
\newblock Cambridge University Press, 2007.

\bibitem{Braun:2008jp}
V.~Braun, T.~Brelidze, M.~R. Douglas, and B.~A. Ovrut, ``{Eigenvalues and
  Eigenfunctions of the Scalar Laplace Operator on Calabi-Yau Manifolds},''
  \href{http://dx.doi.org/10.1088/1126-6708/2008/07/120}{{\em JHEP} {\bf 07}
  (2008)  120}, \href{http://arxiv.org/abs/0805.3689}{{\tt arXiv:0805.3689
  [hep-th]}}.

\bibitem{Ashmore:2020ujw}
A.~Ashmore, ``{Eigenvalues and eigenforms on Calabi{\textendash}Yau
  threefolds},'' \href{http://dx.doi.org/10.1016/j.geomphys.2023.105028}{{\em
  J. Geom. Phys.} {\bf 195} (2024)  105028},
  \href{http://arxiv.org/abs/2011.13929}{{\tt arXiv:2011.13929 [hep-th]}}.

\bibitem{Mirbabayi:2026saz}
M.~Mirbabayi and G.~Villadoro, ``{How Heavy Can Moduli Be?},''
  \href{http://arxiv.org/abs/2603.09817}{{\tt arXiv:2603.09817 [hep-th]}}.

\bibitem{Bertucci:2026osd}
F.~Bertucci, J.~Bonifacio, and K.~Hinterbichler, ``{Moduli Bounds from Spin-2
  Sum Rules},'' \href{http://arxiv.org/abs/2608.05273}{{\tt arXiv:2608.05273
  [hep-th]}}.

\bibitem{Bertucci:2026omk}
F.~Bertucci, ``{Fermionic Kaluza-Klein Reductions},''
  \href{http://arxiv.org/abs/2611.08584}{{\tt arXiv:2611.08584 [hep-th]}}.

\bibitem{Th1984}
T.~Friedrich, ``{Zur Abhängigkeit des Dirac-operators von der
  Spin-Struktur},'' {\em Colloquium Mathematicae} {\bf 48} (1984) no.~1,
  57--62.

\bibitem{diracSpect_book}
N.~Ginoux, \href{http://dx.doi.org/10.1007/978-3-642-01570-0}{{\em {The Dirac
  Spectrum}}}, vol.~1976 of {\em Lecture Notes in Mathematics}.
\newblock Springer Berlin, Heidelberg, 2009.

\bibitem{Higuchi:1986wu}
A.~Higuchi, ``{Symmetric Tensor Spherical Harmonics on the $N$ Sphere and Their
  Application to the De Sitter Group SO($N$,1)},''
  \href{http://dx.doi.org/10.1063/1.527513}{{\em J. Math. Phys.} {\bf 28}
  (1987)  1553}. [Erratum: J.Math.Phys. 43, 6385 (2002)].

\bibitem{Camporesi:1995fb}
R.~Camporesi and A.~Higuchi, ``{On the Eigen functions of the Dirac operator on
  spheres and real hyperbolic spaces},''
  \href{http://dx.doi.org/10.1016/0393-0440(95)00042-9}{{\em J. Geom. Phys.}
  {\bf 20} (1996)  1--18}, \href{http://arxiv.org/abs/gr-qc/9505009}{{\tt
  arXiv:gr-qc/9505009}}.

\bibitem{199669}
C.~B\"ar, ``{The Dirac operator on space forms of positive curvature},''
  \href{http://dx.doi.org/10.2969/jmsj/04810069}{{\em Journal of the
  Mathematical Society of Japan} {\bf 48} (1996) no.~1, 69--83}.

\bibitem{Berger1971}
M.~Berger, P.~Gauduchon, and E.~Mazet, {\em Le Spectre d'une Vari{\'e}t{\'e}
  Riemannienne}.
\newblock Springer Berlin Heidelberg, Berlin, Heidelberg, 1971.

\end{thebibliography}\endgroup

\end{document}